\documentclass[pra,twocolumn,showpacs,showkeys,preprintnumbers,amsmath,amssymb]{revtex4}
\pdfoutput=1

\usepackage{graphicx}
\usepackage{dcolumn}
\usepackage{longtable}
\usepackage{bm}
\usepackage{hyperref}
\usepackage{color}

\newcommand{\ket}[1]{$\left\vert  #1   \right\rangle $}
\newcommand{\ketm}[1]{\left\vert  #1   \right\rangle}

\newcommand{\mem}[3]{\left\langle #1 \left\vert  #2 \right\vert #3
                     \right\rangle}

\newcommand\ion[2]{#1\,{\small\rmfamily\uppercase\expandafter{\romannumeral
#2}}\relax}%
\newcommand\ionsmall[2]{#1\,{\footnotesize\rmfamily\uppercase\expandafter{\romannumeral
#2}}\relax}%

\begin{document}

\preprint{draft version \today}

\title{Electron-ion recombination of Si\,{\normalsize IV} forming
Si\,{\normalsize III}: Storage-ring measurement and
multiconfiguration Dirac-Fock calculations}

\author{E. W. Schmidt}
\author{D. Bernhardt}
\author{A. M\"{u}ller}
\author{S. Schippers}
\affiliation{Institut f\"{u}r Atom- und Molek\"{u}lphysik,
             Justus-Liebig-Universit\"{a}t, Leihgesterner Weg 217,
             35392 Giessen, Germany}
\author{S. Fritzsche}
\affiliation{Institut f\"{u}r Physik, Universit\"{a}t Kassel, Heinrich-Plett
Strasse 40, 34132 Kassel, Germany}
 \altaffiliation{Present address: Gesellschaft f\"{u}r
Schwerionenforschung (GSI), Planckstrasse 1, D-64291 Darmstadt,
Germany}
\author{J. Hoffmann}
\author{A. S. Jaroshevich}
 \altaffiliation{Permanent address, Institute of Semiconductor
Physics, 630090 Novosibirsk, Russia}
\author{C. Krantz}
\author{M. Lestinsky}
 \altaffiliation{Present address: Columbia Astrophysics Laboratory,
Columbia University, 550 W. 120th St., MC 5247 New York, NY 10027,
USA}
\author{D. A. Orlov}
\author{A. Wolf}
\affiliation{Max-Planck-Institut f\"{u}r Kernphysik, Saupfercheckweg 1,
69117 Heidelberg, Germany}
\author{D. Luki\'c}
 \altaffiliation{On leave from the Institute of Physics, 10001
Belgrade, Serbia}
\author{D. W. Savin}
 \affiliation{Columbia Astrophysics Laboratory, Columbia University,
550 W. 120th St., MC 5247 New York, NY 10027, USA}

\date{\today}

\begin{abstract}
The electron-ion recombination rate coefficient for \ionsmall{Si}{4}
forming \ionsmall{Si}{3} was measured at the heavy-ion storage-ring
TSR. The experimental electron-ion collision energy range of
$0$--$186~\text{eV}$ encompassed the $2p^6\,nl\,n'l'$ dielectronic
recombination (DR) resonances associated with $3s\to nl$ core
excitations, $2s\,2p^6\,3s\,nl\,n'l'$ resonances associated with
$2s\to nl$ ($n=3,4$) core excitations, and $2p^5\,3s\,nl\,n'l'$
resonances associated with $2p\to nl$ ($n=3,\ldots,\infty$) core
excitations. The experimental DR results are compared with
theoretical calculations using the multiconfiguration Dirac-Fock
(MCDF) method for DR via the $3s \to 3p\,n'l'$ and $3s \to 3d\,n'l'$
(both $n'=3,\ldots,6$) and $2p^5\,3s\,3l\,n'l'$ ($n'=3,4$) capture
channels. Finally, the experimental and theoretical plasma DR rate
coefficients for \ionsmall{Si}{4} forming \ionsmall{Si}{3} are
derived and compared with previously available results.
\end{abstract}



\maketitle

\section{Introduction}\label{sec::intro}
Spectroscopic \cite{Lukic2007a} observations of absorption lines in
the intergalactic medium (IGM) can be used to study the origin of
large-scale structure in the universe, the history of star and galaxy
formation, the metagalactic radiation field, and the chemical
evolution of the IGM \citep{Savin2000b,Levshakov2002,Aguirre2004a}.
Observations of lines from \ion{C}{4}, \ion{N}{5}, \ion{O}{6}, and
\ion{Si}{4} are routinely employed for these studies. These
observations provide important constraints for IGM studies when
coupled with calculations of the ionization balance using codes such
as CLOUDY \citep{Ferland1998}. However, the accuracy with which one
can infer the properties of the IGM is limited by uncertainties in
the underlying atomic data.

Of particular importance are reliable electron-ion recombination
data for the process known as dielectronic recombination (DR). This
is the dominant recombination process for most atomic ions under IGM
conditions. Recently, \citet{Savin2000b} has investigated the
importance of DR for \ion{C}{4}, \ion{N}{5}, \ion{O}{6}, and
\ion{Si}{4}. (Here, the convention of identifying each ion by its
charge state before the recombination process is used.) His work has
shown that uncertainties in the DR data for these four ions limit
our ability to constrain the metagalactic radiation field and the
initial mass function for the earliest generations of stars.

In the last several years, a series of measurements has been carried
out to produce accurate DR data for \ion{C}{4}
\citep{Schippers2001c}, \ion{N}{5} \citep{Boehm2005a}, \ion{O}{6}
\citep{Boehm2003a}, and sodium-like \ion{Si}{4} which will be
presented in this work. Other experimental results for DR rate
coefficients of Na-like ions have been published for \ion{Si}{4}
\citep{Orban2006a}, for \ion{Fe}{16} \citep{Linkemann1995a}  and for
\ion{Ni}{18} \citep{Fogle2003b}. Recently detailed calculations of
the low energy DR resonance structure have been performed for
\ion{Si}{4} \citep{Orban2007a} within the framework of relativistic
many-body perturbation theory (RMBPT).

In the present work, experimental results for the \ion{Si}{4}
recombination rate coefficient are presented. These were obtained
employing the electron-ion merged-beams method at a heavy-ion storage
ring. \citet{Orban2006a} measured the \ion{Si}{4} recombination rate
coefficient in the electron-ion collision energy range
$0$--$20~\text{eV}$ that comprises DR resonances associated with
$3s\to 3p$ and $3s\to 3d$ ($\Delta N=0$) core excitations. Here an
extended energy range of up to $186~\text{eV}$ was experimentally
investigated. This additional range includes DR resonances associated
with $3s\to nl$ ($n\geq 4$), $2p\to nl$ ($n=3,\ldots,\infty$)
($\Delta N =1,2,\ldots$) and $2s\to nl$ ($n=3,4$) ($\Delta N=1,2$)
core excitations. The corresponding excitation energies are listed in
Table~\ref{table::excenergy}, except those for $2s\to nl$ excitation.
The present experimental \ion{Si}{4} merged-beams recombination rate
coefficient thus benchmarks theory for light, low charged sodium-like
ions over a wide range of energies.
\begin{table}[htb]
\caption{\label{table::excenergy}Excitation energies $E_\text{exc}$
for the excitation of \ion{Si}{4} $2s^2\,2p^6\,3s\,^2S_{1/2}$ ground
state to $2p^6\,nl$ and to $2p^5\,3s\,nl$ (both $n=3,4$) states that
are relevant in the present work. For the derivation of the
$2p^5\,3s\,nl$ energies, results from experimental Auger spectroscopy
\citep{Schneider1995} were added to the \ion{Si}{4} ionization energy
of $45.14179~\text{eV}$ \citep{Ralchenko2006a}.}
\begin{ruledtabular}
\begin{tabular}{ldc}
 excited state & \multicolumn{1}{c}{$E_\text{exc}$ (eV)} &
\multicolumn{1}{c}{Reference}\\
\hline
\vspace{-2ex}\\
$2p^6\,3p                          \,^2\text{P}_{1/2}$    &  8.839
&\citep{Toresson1960}\\
$2p^6\,3p                          \,^2\text{P}_{3/2}$    &  8.896
&\citep{Toresson1960}\\
$2p^6\,3d                          \,^2\text{D}_{5/2}$    & 19.884
&\citep{Ralchenko2006a}\\
$2p^6\,3d                          \,^2\text{D}_{3/2}$    & 19.884
&\citep{Ralchenko2006a}\\
$2p^6\,4s                          \,^2\text{S}_{1/2}$    & 24.050
&\citep{Ralchenko2006a}\\
$2p^6\,4p                          \,^2\text{P}_{1/2}$    & 27.062
&\citep{Ralchenko2006a}\\
$2p^6\,4p                          \,^2\text{P}_{3/2}$    & 27.082
&\citep{Ralchenko2006a}\\
$2p^6\,4d                          \,^2\text{D}_{5/2}$    & 30.997
&\citep{Ralchenko2006a}\\
$2p^6\,4d                          \,^2\text{D}_{3/2}$    & 30.997
&\citep{Ralchenko2006a}\\
$2p^6\,4f                          \,^2\text{F}_{5/2}$    & 31.508
&\citep{Ralchenko2006a}\\
$2p^6\,4f                          \,^2\text{F}_{7/2}$    & 31.508
&\citep{Ralchenko2006a}\\
$2p^5\,3s^2                        \,^2\text{P}_{3/2}$    & 99.06
&\citep{Schneider1995}\\
$2p^5\,3s^2                        \,^2\text{P}_{1/2}$    & 99.68
&\citep{Schneider1995}\\
$2p^5\,3s  \,3p\,(^3\text{P})      \,^4\text{S}_{3/2}$    &104.8
&\citep{Schneider1995}\\
$2p^5\,3s  \,3p\,(^3\text{P})      \,^4\text{D}      $    &106.2
&\citep{Schneider1995}\\
$2p^5\,3s  \,3p\,(^3\text{P})      \,^4\text{P}      $    &106.9
&\citep{Schneider1995}\\
$2p^5\,3s  \,3p\,(^3\text{P})      \,^2\text{D}      $    &107.5
&\citep{Schneider1995}\\
$2p^5\,3s  \,3p\,(^3\text{P})      \,^2\text{S}_{1/2}$    &108.4
&\citep{Schneider1995}\\
$2p^5\,3s  \,3p\,(^1\text{P})      \,^2\text{D}_{5/2}$    &110.9
&\citep{Schneider1995}\\
$2p^5\,3s  \,3p\,(^1\text{P})      \,^2\text{D}_{3/2}$    &111.5
&\citep{Schneider1995}\\
$2p^5\,3s  \,3p\,(^1\text{P})      \,^2\text{P}      $    &112.0
&\citep{Schneider1995}\\
$2p^5\,3s  \,3p\,(^1\text{P})      \,^2\text{S}_{1/2}$    &112.7
&\citep{Schneider1995}\\
$2p^5\,3s  \,3d\,(^3\text{D})      \,^4\text{P}_{3/2}$    &119.2
&\citep{Schneider1995}\\
$2p^5\,3s  \,3d\,(^3\text{D})      \,^4\text{F}      $    &119.9
&\citep{Schneider1995}\\
$2p^5\,3s  \,3d\,(^3\text{D})      \,^4\text{D}_{7/2}$    &121.1
&\citep{Schneider1995}\\
$2p^5\,3s  \,3d\,(^1\text{D})      \,^2\text{F}_{7/2}$    &124.1
&\citep{Schneider1995}\\
$2p^5\,3s  \,3d\,(^1\text{D})      \,^2\text{D}_{3/2}$    &124.6
&\citep{Schneider1995}\\
$2p^5\,3s  \,3d\,(^1\text{D})      \,^2\text{D}_{1/2}$    &125.1
&\citep{Schneider1995}\\
$2p^5\,3s      \,(^3\text{P})  \,4s\,^4\text{P}_{3/2}$    &125.8
&\citep{Schneider1995}\\
$2p^5\,3s      \,(^3\text{P})  \,4s\,^2\text{P}      $    &126.3
&\citep{Schneider1995}\\
$2p^5\,3s      \,(^1\text{P})  \,4s\,^2\text{P}_{3/2}$    &126.9
&\citep{Schneider1995}\\
$2p^5\,3s  \,4p\,(^3\text{P})      \,^4\text{S}      $    &128.4
&\citep{Schneider1995}\\
$2p^5\,3s  \,4p\,(^3\text{P})      \,^2\text{P}      $    &129.5
&\citep{Schneider1995}\\
$2p^5\,3s  \,4p\,(^1\text{P})      \,^2\text{P}      $    &131.0
&\citep{Schneider1995}\\
$2p^5\,3s  \,4d\,(^3\text{D})      \,^4\text{F}      $    &133.3
&\citep{Schneider1995}\\
$2p^5\,3s  \,4d\,(^3\text{D})      \,^2\text{P}      $    &134.7
&\citep{Schneider1995}\\
\end{tabular}
\end{ruledtabular}
\end{table}

In this work, the experimental results are compared to theoretical
results using the multiconfiguration Dirac-Fock (MCDF) method, in
particular at low electron-ion collision energies. At these energies,
the calculation of accurate resonance positions and strengths is
extremely critical for the derivation of reliable plasma DR rate
coefficients. Therefore, the present work is partly an investigation
into the capabilities of the MCDF method for calculating accurate DR
resonance energies and strengths. The MCDF method is an ab-initio
method that is applicable to target atoms and ions with an arbitrary
shell structure. In contrast, the RMBPT is currently limited to quasi
one-electron target ions, owing to its inherent complexity. For such
systems, however, it usually yields more accurate results than any
other theoretical method. Below, a detailed comparison between the
present MCDF results and the RMPBT results of \citet{Orban2007a} for
DR of sodiumlike \ion{Si}{4} will be presented.

The present paper is organized as follows. The theoretical procedure
is outlined in Section~\ref{sec::theory}. In
Section~\ref{sec::experiment} the experimental procedure is
described. Experimental and theoretical results are presented and
compared in Section~\ref{sec::mergedbeamcoefficients}. The
\ion{Si}{4} to \ion{Si}{3} experimental and theoretical DR rate
coefficients in a Maxwellian plasma are derived and compared to
recent theoretical and experimental results in
Section~\ref{subsec::plasma}. Conclusions will be presented in
Section~\ref{sec::conclusions}.

\section{Theory and implementation}\label{sec::theory}
In most computations, DR of an $\mathcal{N}$-electron target ion in
the initial state \ket{i} is handled as a two--step process in which
first an electron is captured resonantly from the continuum forming
an $(\mathcal{N}+1)$-electron state \ket{d} with the captured
electron and one of the target core electrons both now in an excited
level. In a second step, this doubly excited state either decays
radiatively by the emission of one or more photons to some final
state \ket{f} which lies below the ionization threshold of the ion,
or it returns by Auger electron emission to the initial ionization
stage of the target. If interference between the radiative and
nonradiative capture of the electron in the field of the target is
negligible, the integrated DR cross section for an isolated
resonance, the so--called resonance strength, can be expressed in
terms of the Auger and radiative rates of the intermediate state
\ket{d} by
\begin{eqnarray}
\label{eq::resonance-strength}
   S (i \to d \to f) & \equiv & \int_{-\infty}^{\infty} \,
\sigma_{\rm\, DR} (E) \, dE\nonumber\\*
   &=& \frac{2\pi^2 \, \hbar}{k_i^2} \:\frac{A_a (i \to d) \, A_r (d
\to f)}{\Gamma_d},
\end{eqnarray}
where $k_i$ denotes the wave number and $E$ the energy of the
incident electron, $\Gamma_d$ the width of the doubly--excited state,
and $A_a (i \to d)$ is the rate for the dielectronic capture from the
initial into the doubly excited state. Using the principle of
detailed balance, the dielectronic capture rate $A_a (i \to d)$ is
equal to $[g_d /g_i \: A_a (d \to i)]$ where $A_a (d \to i)$ is the
Auger rate, and $g_i$ and $g_d$ are the statistical factors of the
initial and intermediate states, respectively. $A_r (d \to f)$ refers
to the rate for the radiative stabilization to the state \ket{f}. The
strength as defined by Equation~(\ref{eq::resonance-strength}) refers
to the area under the DR cross section curve. Hence, it is usually
given in units of $\text{cm}^2\,\text{eV}$. The total width
$\Gamma_d$ is determined by all possible decay channels of the
resonant state \ket{d} and is given, in first--order perturbation
theory, by
\begin{eqnarray}
\label{eq::total-width}
   \Gamma_{d} & = &
   \hbar \: \left( \sum_{j} A_a (d \to j) \;+\;
                   \sum_{f'} A_r (d \to f') \right),
\end{eqnarray}
where the sums are over all the individual Auger and radiative rates
(widths) of the intermediate state $\ketm{d}$. The use of the
resonance strength $S (i \to d \to f)$ is appropriate for resonances
whose width $\Gamma_d$ is small compared to its energy position
$E_\text{res} = E_d - E_i$ where  $E_d$ and $E_i$ are the energy of
the doubly excited state and initial state of the target ion,
respectively, and if the energy dependent DR cross section has a
Lorentzian profile \citep{Tokman2002}
\begin{eqnarray}
\label{eq::lorentzian-profile}
   \sigma_{\rm\, DR} (E) & = & \frac{S}{\pi} \,
   \frac{\Gamma_d / 2}{(E_\text{res} - E)^2 + \Gamma_{d}^2/4}.
\end{eqnarray}
Owing to the energy of the incident electron [cf.\ the $1 / k_i^2$
factor in $S$], the resonance strength increases rapidly towards the
DR threshold and makes the process sensitive to low kinetic energies
of the incoming electrons.

At storage rings, the radiative stabilization is often not observed
explicitly and, hence in Equation~(\ref{eq::resonance-strength}),
the radiative rate for the individual transition $d \to f$ has to be
replaced with
\begin{eqnarray*}
   A_r (d \to f) \;\longrightarrow \, \sum_{f''}  A_r (d \to f''),
\end{eqnarray*}
where the summation extends over all states \ket{f''} below the
ionization threshold that can be reached by radiative transition from
the doubly excited state \ket{d}. As seen from
Equation~(\ref{eq::resonance-strength}), the observed DR strength for
a given resonance is proportional to the capture rate $A_a (i \to d)$
and the total radiative rate. For most light and medium elements,
moreover, the magnitude of the resonance strengths is often
determined by the radiative decay since the emission of photons from
the doubly excited state is then much slower than its autoionization
and, hence, $ A_a\,A_r / (A_a + A_r) \,\approx\, A_r$.

The doubly excited state \ket{d} is often one out of a large number
of highly correlated states embedded in the continuum of the target
ion. For these resonances, special care has to be taken in
calculating both the individual as well as total rates accurately. To
describe the ground and excited state of multiply charged ions, the
MCDF method has been found to be a versatile tool for the computation
of the many--electron energies and decay properties, especially if
inner--shell electrons or several open shells are involved in the
computations \citep{Grant1988,Fritzsche2002}. In the MCDF method, an
atomic state is approximated by a linear combination of so--called
configuration state functions (CSF) of the same symmetry
\begin{eqnarray}
\label{eq::ASFansatz}
  \psi_{\alpha}(PJ)=\sum_{r=1}^{n_c} \, c_r(\alpha) \,
\ketm{\gamma_rPJ},
\end{eqnarray}
where $n_c$  is the number of CSF, $\{ c_r(\alpha) \}$ denotes the
representation of the atomic state in this CSF basis, $P$ is the
parity, $J$ is the total angular momentum, and $\gamma_r$ is a set of
quantum numbers for a unique specifying of the many-electron basis
states. In most standard computations, the CSF are constructed as
antisymmetrized products of a common set of orthonormal orbitals and
are optimized on the basis of the Dirac--Coulomb Hamiltonian. Further
relativistic contributions to the representation $\{ c_r(\alpha) \}$
of the atomic states are then added, owing to the given requirements,
by diagonalizing the Dirac--Coulomb--Breit Hamiltonian matrix in
first--order perturbation theory. For multiply charged ions, an
estimate of the dominant QED contributions (i.e., the self--energy
and vacuum polarization of the electronic cloud) might be taken into
account by means of scaled hydrogenic values. But QED plays a
negligible role for Si$^{3+}$ ions in the present analysis.

Obviously, the calculation of the radiative and nonradiative decay
rates of the intermediate resonances \ket{d} is central to the
computation of any DR spectrum which is to be compared with
experiment. In the MCDF model, both the radiative as well as the
Auger matrix elements are derived from computation of the
corresponding interaction matrix within the CSF basis using the
expansion (\ref{eq::ASFansatz}). To determine the rates, the wave
functions from the GRASP92 \citep{Parpia1996} and RATIP
\citep{Fritzsche2001} codes have been applied in the present work
which allows one to incorporate both the dominant relativistic and
correlation effects on the same footings. However, since the
computation of the transition probabilities has been considered at
many places elsewhere \citep{Parpia1996,Fritzsche2002}, we shall
mention here only that the radiative rates
\begin{eqnarray}
\label{eq::Ar-rate}
  A_r (d \to f) & = & \frac{4}{3\,c^2} \,\frac{\omega_{d \to
f}^3}{2J_d +1} \times\\*
  &&\sum_{L\pi} \, \vert
  \mem{\psi_f (J_f P_f)}{H_\gamma (\pi L)}{\psi_d (J_d P_d)}
\vert^2,\nonumber
\end{eqnarray}
where $\omega_{d \to f}$ is the emitted photon's frequency, are
associated with the reduced matrix elements of the interaction
$H_\gamma (\pi L)$ of the atomic electrons with the multipole
components of the radiation field \citep{Fritzsche2005}, where
$\pi=0$ and $\pi=1$ refer to electric and magnetic multipoles,
respectively, with the multipolarity $L$. For light and medium
elements, it is of course sufficient to include the electric--dipole
($L=1$, $\pi=1$) decay while the contributions from higher multipoles
to the radiative stabilization remain negligible. To obtain the total
radiative rate, i.e., to include the summation over all lower states
\ket{f'} in Equation~(\ref{eq::total-width}), often a large number of
transition rates have to be compiled with similar accuracy. For this
reason the computation of the DR spectrum at higher energies may
become difficult. In the RATIP program, the transition probabilities
are computed by the REOS component \citep{Fritzsche2000a} including,
if appropriate, the rearrangement of the electron density in course
of the decay. The explicit consideration of the electron density's
rearrangement was not included here, because this would require
separate calculations for the intermediate resonance states and the
final states of the radiative stabilization. In the present
calculations, however, both sets of states were always treated
together in order to keep the computations feasible.

Unlike the treatment of radiative stabilization, calculation of the
Auger rates requires the coupling of the bound--state electrons (of
the target ion) to the electron continuum. If, for the sake of
simplicity, we neglect the interaction between different Auger
channels (i.e.\ \emph{within the continuum}), the Auger rates are
given by
\begin{eqnarray}
\label{eq::Aa-rate}
  &&A_a (d \to j) = 2\pi \times\\*
   &&\sum_{\kappa_c}
  \vert \mem{(\psi_j (J_j P_j),\, \epsilon \kappa_c) \, J_d P_d
           }{\,H- E_d\,}{\psi_d (J_d P_d)} \vert^2, \nonumber
\end{eqnarray}
where $H$ is the atomic Hamiltonian, $E_d$ the total energy of the
doubly excited state and $\epsilon = E_d - E_j$ the kinetic energy of
the emitted electron. In Equation~(\ref{eq::Aa-rate}), the summation
over $\kappa_c = \pm (j_c + 1/2)$ for $l = j_c \pm 1/2$ , where $l$
and $j_c$ are the angular momentum and total angular momentum,
respectively, of the incoming or outgoing electron, extends over all
partial waves  of the outgoing electron which can be coupled to the
target ion state \ket{\psi_j (J_j,P_j)} with the condition to
conserve the total angular momentum $J_d$ and parity $P_d$ of the
intermediate state. If, moreover, a common set of \emph{orthonormal}
orbitals is used for the representation of the intermediate state
\ket{d} and the final ionic states \ket{j}, then the operator $(H -
E) \;\approx\, V$ can be replaced by the electron--electron
interaction operator. For most light and medium elements, it is again
sufficient to include the instantaneous Coulomb repulsion between the
electrons but to omit the relativistic Breit contributions as they
were found important only for the Auger emission of highly--charged
ions \citep{Zimmerer1990,Chen1990a,Badnell1991,Fritzsche1991}.

The restriction of the electron--electron interaction in the
computation of the matrix elements in Equation~(\ref{eq::Aa-rate}) is
common practice, even though the orbital functions for the doubly
excited state \ket{d} and the final ionic state \ket{j} are not quite
orthogonal to each other. This treatment has been implemented
therefore also in the AUGER component of the RATIP program, in which
the continuum spinors are solved within a spherical but
level--dependent potential of the final ion  (the so--called
\emph{optimal level} scheme in the GRASP92 program). This scheme also
includes the exchange interaction of the emitted electron with the
bound--state density. Often, the number of the possible scattering
states \ket{(\psi_j (J_j P_j),\, \epsilon \kappa_c) \, J_d P_d } of a
system increases rapidly with the number of intermediate ion states
as the free electrons may couple in quite different ways to the
bound--state electrons. For further details on the computations of
the Auger matrix elements and relative intensities, we refer the
reader to \citep{Fritzsche2002} and \citep{Fritzsche1992}.

Apart from the individual rates, any helpful calculation of DR
spectra critically depends on the proper control and handling of the
various decay branches which, at least in principle, should be
treated on the same basis in order to obtain a consistent spectrum.
This need for an efficient handling of the decay channels concerns
those open-shell configurations in which one-particle excitations
with large principal quantum numbers `mix into' the low-energy part
of the DR spectrum. Such a mixing occurs especially for the
excitation of electrons beneath the valence shell whose energies are
comparable with the valence excitation into Rydberg states.
Therefore, in order to facilitate this handling of the various decay
branches, a new component (DIEREC) has been developed recently in the
framework of the RATIP program. This component now supports both, the
computation of individual $S (i \to d \to f)$ and total resonance
strengths $S (i \to d)$ and enables us to \emph{simulate} the
low--energy DR spectra as observed by experiment. Internally, of
course, this new component makes use of the REOS and AUGER components
from above and allows us, if necessary, to incorporate also higher
multipoles in the computation of the (total) radiative rates. Because
of the finite--difference method of the GRASP92 and RATIP programs
and owing to the associated size of the configuration expansions,
however, excitation to states with principal quantum numbers $n
\,\gtrsim\, 8$ can often not be treated by the code explicitly.
Atomic structure codes, such as, e.g., AUTOSTRUCTURE
\citep{Badnell1986}, that are geared towards the generation of DR
rate coefficients for plasma physical applications, treat high-$n$
electrons within a hydrogenic approximation. This has not yet been
implemented in the RATIP program suite.

\section{Experiment}\label{sec::experiment}
The experiment was performed at the heavy-ion storage ring TSR of
the Max-Planck-Institut f\"{u}r Kernphysik (MPI-K) in Heidelberg,
Germany. Details of the merged-beams technique using the MPI-K
electron cooler have been described previously
\citep{Kilgus1992,Lampert1996,Pastuszka1996,Schippers2001c,Mueller1997c}.

In the present experiment a beam of $^{28}$Si$^{3+}$ was provided by
the tandem accelerator of the MPI-K linear accelerator facility at an
energy of about $1.1~\text{MeV/u}$. The ion beam was injected into
the storage ring where it was collinearly overlapped with the cooler
electron beam. For efficient electron cooling, the electron velocity
has to be close to the ion velocity in the ring. To fulfill this
velocity-matching condition, the laboratory energy of the electrons
was set to the `cooling' energy $E_c\approx 593~\text{eV}$. The beam
current was accumulated by multi-turn injection and `ecool stacking'
\citep{grieser1991}. Ion currents of typically $10$--$50~\mu\text{A}$
were reached.

In contrast to previous experiments, where the electron beam of the
cooler was also used as an electron target for recombination
experiments, in the present experiment a newly installed separate
electron beam \citep{Sprenger2004a} was used. This additional
electron beam is hereafter denoted as the electron target. As in the
electron cooler, the electron beam of the electron target is also
guided by a magnetic field and overlaps the ion beam in a straight
section of $\approx 1.5~\text{m}$ length.

Conceptually, the experimental procedures for measuring
recombination rate coefficients with the electron target are the
same as those applied previously with the electron cooler. However,
there are advantages when a separate electron target is employed for
recombination measurements. First, the electron cooler can be used
continuously for the cooling of the ion beam. Thus, the low velocity
and spatial spread of the ion beam is maintained at all times.
Second, the electron target was specifically designed for providing
an electron beam with a very low initial energy spread
\citep{Sprenger2004a}. Both advantages yield a higher experimental
resolution in the present measurement as compared to previous
measurements using only the electron cooler.

After injection into the storage ring, the ions were cooled for a few
seconds before the recombination measurements started. Recombined
ions were separated from the circulating beam in the first dipole
magnet downstream of the electron target and counted by a single
particle scintillation detector with nearly $100\%$ efficiency.
During the recombination measurement, the electron energy of the
electron target beam was alternatingly chopped between measurement
($E_m$) and reference ($E_r$) energies by switching the acceleration
voltage for the target electron beam accordingly. At the same time
the electron cooler was held constant at cooling energy. The
reference measurement was made to determine detector background.
Therefore, the reference energy $E_r$ was chosen in a range of the
spectrum where no DR resonances occurred. The measurement and
reference interval of data collection were $10~\text{ms}$ each.
Between the voltage jump and the data collection interval there was a
settling time of $5~\text{ms}$ to allow the power supplies to reach
the preset values. The data were collected in overlapping data sets
ranging from laboratory energies $E_m=561~\text{eV}$ to
$E_m=1420~\text{eV}$. The chopping pattern was
$E_m^{\nu}$--$E_r$--$E_m^{\nu+1}$--$E_r$ with $\nu=1,2,3\ldots$. With
each step $\nu$ in the chopping pattern $E_m$ was changed by
$0.15~\text{eV}$ in the laboratory frame whereas $E_r$ was kept
fixed. The merged-beams rate coefficient is derived from the
background subtracted recombination count rate using
\citep{Kilgus1992,Poth1988a}
\begin{equation}\label{eq::alpha}
\alpha(E_m)=
 \frac{[R(E_m)-R(E_m,\,E_r)]}
      {(1-\beta_\text{i}\beta_\text{e})\eta
n_\text{e}(E_m)N_\text{i}L/C}
 +\alpha(E_r)\frac{n_\text{e}(E_r)}
 {n_\text{e}(E_m)},
\end{equation}
where $R$ is the recombination count rate, $\eta$ is the detection
efficiency, $n_\text{e}$ is the electron density in the interaction
region, $N_\text{i}$ is the number of ions in the ring and
$C=55.4~\text{m}$ is the circumference of the TSR. In a first
approach the nominal length $L$ of the electron-ion overlap region in
the electron target section is set to $L=1.476~\text{m}$ which is the
length of the solenoid providing the axial magnetic field along the
interaction region. For the detailed analysis we applied a toroid
correction \citep{Lampert1996} that accounts for the contribution of
the merging and de-merging sections in the toroidal magnetic fields
of the electron target. The ion velocity and the electron velocity
are $v_\text{i}=\beta_\text{i}c$ and $v_\text{e}=\beta_\text{e}c$,
respectively, where $c$ is the speed of light. Under the present
conditions, the factor $1/(1-\beta_\text{i}\beta_\text{e})$ deviates
by less than $0.4\%$ from unity. The second term in
Equation~(\ref{eq::alpha}) is a small correction that re-adds the
recombination rate coefficient $\alpha(E_r)$ at the reference point.
This contribution was calculated using a hydrogenic formula
\citep{Schippers2001c} for nonresonant radiative recombination (RR).
The systematic experimental uncertainty of the merged-beams rate
coefficient is estimated to be $\pm18\%$ at $1\sigma$ confidence
level. This uncertainty stems mostly from the ion current measurement
($\pm15\%$ including also the error of the toroid correction)
\citep{Lampert1996} and the determination of the electron density
($\pm10\%$).

The experimental electron energy distribution is best described as a
flattened Maxwellian distribution which is characterized by the
longitudinal and transverse temperatures $T_\parallel$ and $T_\perp$.
The experimental energy resolution is approximately given by
$\Delta{E}=[(\ln(2)\, k_\text{B}T_\perp)^2 +
16\ln(2)\,{E}k_\text{B}T_\parallel]^{1/2}$, where $E$ denotes the
relative electron-ion energy. For the reduction of $T_\perp$ the
target electron beam was adiabatically expanded \citep{Pastuszka1996}
by factors of up to $28$.

For the measurements we used two different cathodes, a thermionic
cathode and a photocathode \citep{Orlov2004a}. The electron current
produced by the thermionic cathode was $4~\text{mA}$ and that of the
photocathode $0.5~\text{mA}$.

The cryogenic photocathode provides electrons with a laboratory
energy spread of about $10~\text{meV}$ \cite{Orlov2004a}. The
expansion factor was $20$. The electron beam temperatures were
estimated by fitting a simulated (Sec.~\ref{subsec::expresults})
merged-beams recombination rate-coefficient comprising contributions
by DR and RR to the measured spectrum. A transverse temperature of
$k_\text{B}T_\perp\approx0.9~\text{meV}$ and a longitudinal
temperature of $k_\text{B}T_\parallel\approx 35~\mu\text{eV}$ were
found. With the above given temperature values, the experimental
energy spread in the center of mass frame thus amounts to
 $\Delta{E}=6~\text{meV}$ at ${E}=0.1~\text{eV}$,
 $\Delta{E}=20~\text{meV}$ at ${E}=1~\text{eV}$,
 $\Delta{E}=62~\text{meV}$ at ${E}=10~\text{eV}$,
 and
 $\Delta{E}=197~\text{meV}$ at ${E}=100~\text{eV}$.

The temperature of the thermionic cathode is typically about
$1300~\text{K}$. The expansion factor was 28. The temperatures were
again derived by fitting a simulated spectrum to the measured one. A
transverse temperature of $k_\text{B}T_\perp\approx4.4~\text{meV}$
and a longitudinal temperature of $k_\text{B}T_\parallel\approx
100~\mu\text{eV}$ were determined. With these temperatures, the
experimental energy spread thus amounts to
 $\Delta{E}=11~\text{meV}$ at ${E}=0.1~\text{eV}$,
 $\Delta{E}=33~\text{meV}$ at ${E}=1~\text{eV}$,
 $\Delta{E}=105~\text{meV}$ at ${E}=10~\text{eV}$,
 and
 $\Delta{E}=333~\text{meV}$ at ${E}=100~\text{eV}$.

\section{Merged-beams rate
coefficients}\label{sec::mergedbeamcoefficients}
\subsection{Experimental results}\label{subsec::expresults}
\begin{figure}[htb]
\includegraphics[width=\columnwidth]{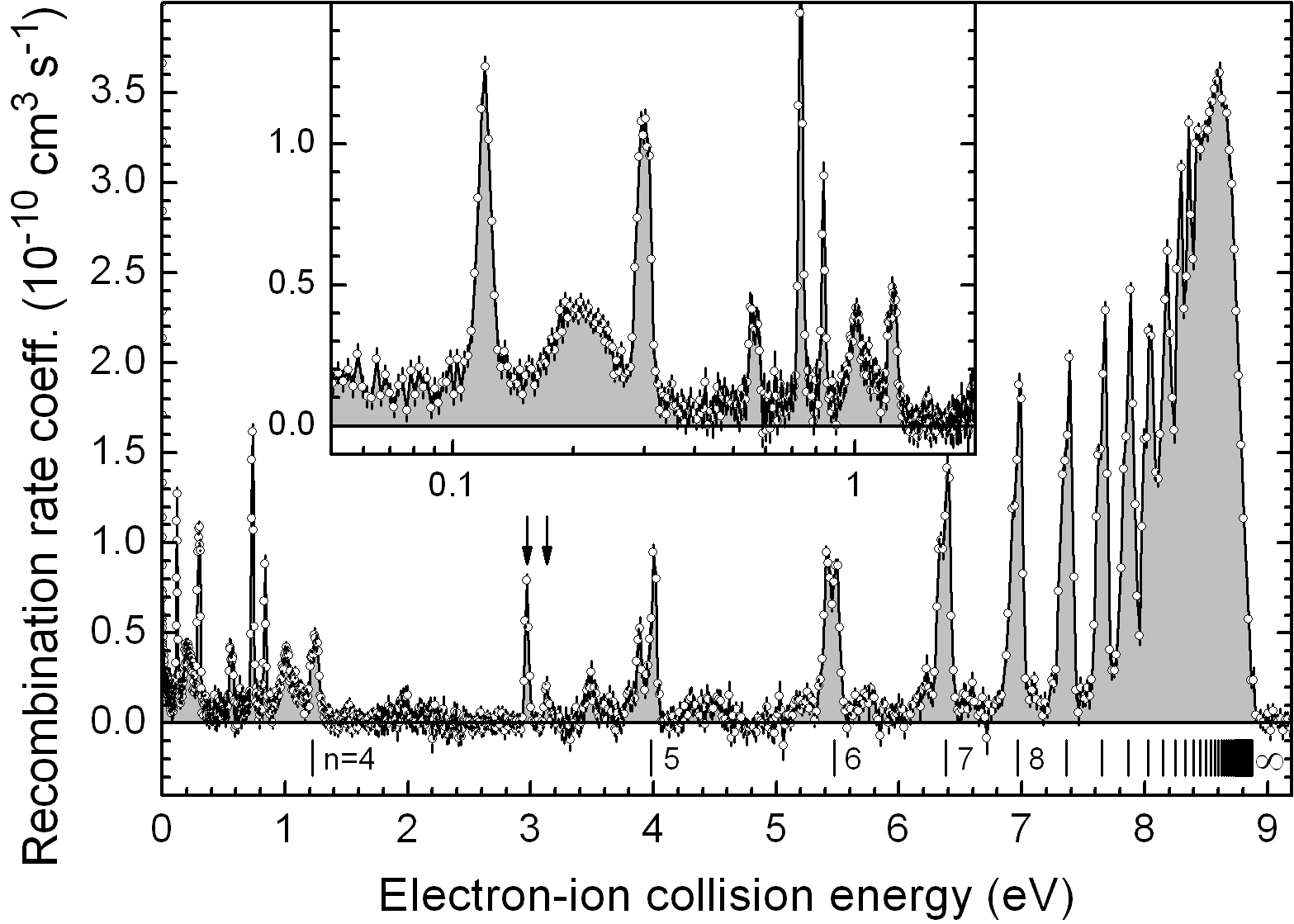}
\caption{Measured \ionsmall{Si}{4} to \ionsmall{Si}{3} merged-beams
electron-ion recombination rate coefficient in the energy range
dominated by DR resonances associated  with $3s\to 3p$ core
excitations. The vertical bars below the spectrum denote the
$2p^6\,3p\,n'l'$ DR resonance positions as expected on the basis of
the hydrogenic Rydberg formula (Eq.~\ref{eq::rydberg}). Note that
resonances up to $n=15$ are individually resolved.  The vertical
arrows around $3~\text{eV}$ denote the positions of the
$2p^6\,3d^2\,^3F$ and $2p^6\,3d^2\,^1G$ doubly excited states. The
inset shows the recombination rate coefficient at energies below
$2~\text{eV}$, where it is dominated by $2p^6\,3p\,4l'$ resonances.}
\label{fig::DN0sp}
\end{figure}
Figure~\ref{fig::DN0sp} shows the measured merged-beams recombination
rate coefficient in the energy range $0$--$9.2~\text{eV}$. For this
measurement the photocathode was used. In this range all DR
resonances associated with $3s\to 3p$ core excitations appear. The
Rydberg series of $2p^6\,3p\,n'l'$ DR resonances converges to its
series limit at $E_{\infty}=8.877~\text{eV}$ \citep{Toresson1960}.
This value is the weighted average of the $3p_{1/2}$ and $3p_{3/2}$
limits. The $3p$ fine structure splitting of $0.057169~\text{eV}$ is
not resolved. Additionally, the $2p^6\,3d\,3l$ and $2p^6\,3d\,4s$
resonances associated with the $3s\to3d$ excitation are expected to
appear at these low energies. As pointed out by \citet{Orban2007a},
in this range the most notable $3s\to3d$ contributions are due to
$2p^6\,3d^2\,^3F$ and $2p^6\,3d^2\,^1G$ doubly excited states whose
energy positions are determined in our experimental data to $2.97$
and $3.13~\text{eV}$, respectively.

The experimental energy scale was fine-tuned by multiplying the
nominal electron-ion collision energies with an energy independent
factor, so that so that the positions of the $2p^6\,3p\,n'l'$ ($7\leq
n\leq10$) resonances matched their calculated positions. For the
high-$n$ values the position of the Rydberg resonances can be
estimated from the hydrogenic Rydberg formula
\begin{equation}\label{eq::rydberg}
E_n=E_{\infty}-R\,\frac{q^2}{n^2},
\end{equation}
with the Rydberg constant $R$ and the charge state $q=3$ of the
initial \ion{Si}{4} ion. The calibration factor for the energy axis
deviated from unity by less than $1\%$.
\begin{figure}[htb]
\includegraphics[width=\columnwidth]{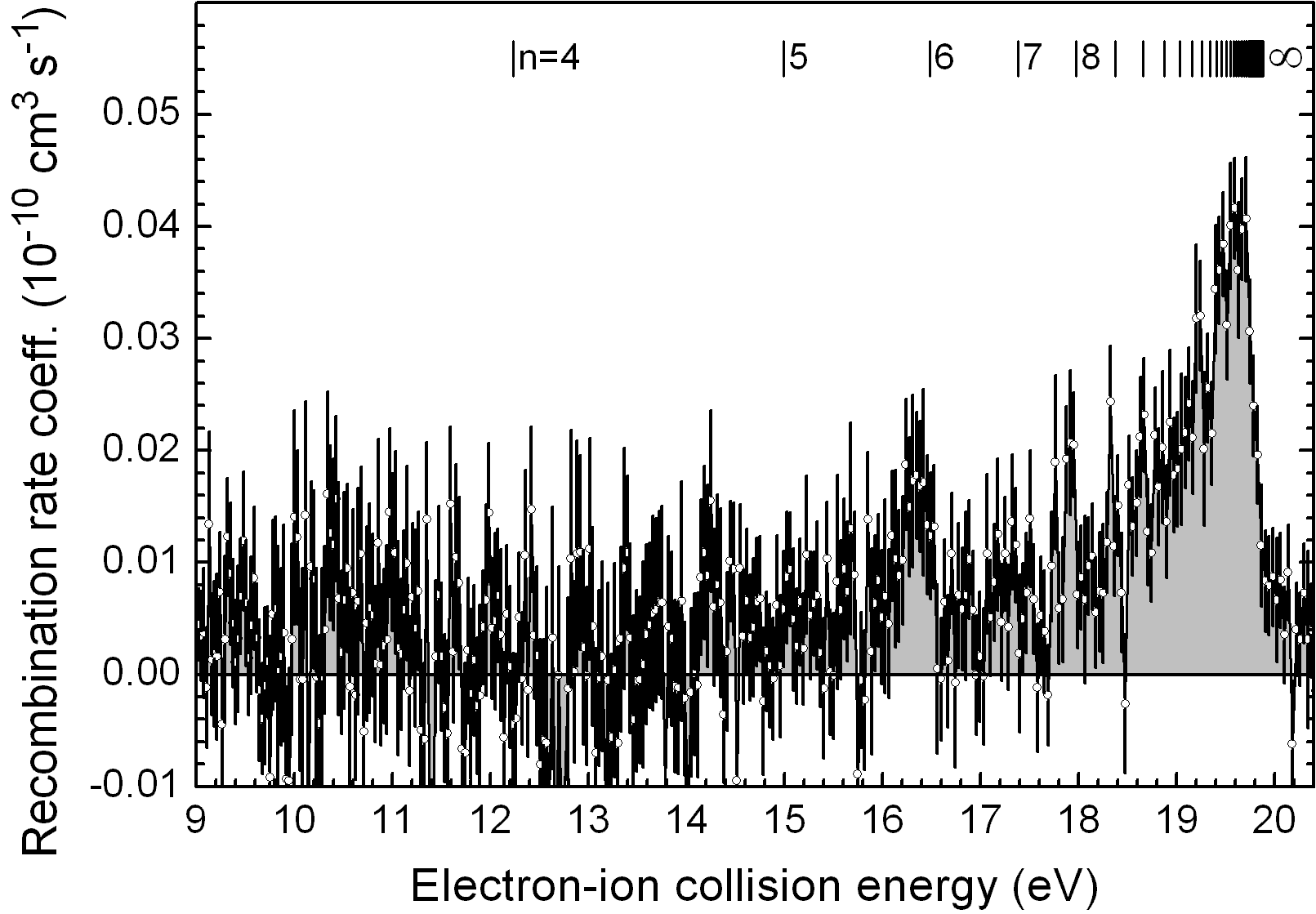}
\caption{Measured \ionsmall{Si}{4} to \ionsmall{Si}{3} merged-beams
electron-ion recombination rate coefficient in the energy range of
DR resonances associated with $3s\to 3d$ core excitations. The
vertical bars denote the $2p^6\,3d\,n'l'$ DR resonance positions as
expected on the basis of the hydrogenic Rydberg formula
[Eq.~(\ref{eq::rydberg})]. The error bars on each data point show
the
statistical uncertainties.} \label{fig::DN0sd}
\end{figure}

The much less intense $2p^6\,3d\,n'l'$ resonances with $n'\ge4$
associated with $3s\to 3d$ core excitations are expected to occur in
the energy range $9$--$20~\text{eV}$. The recombination rate
coefficient measured in this energy range, using the thermionic
cathode, is shown in Figure~\ref{fig::DN0sd}. Individual resonances
of the associated Rydberg series are barely visible with rather large
statistical uncertainties. Nevertheless the $2p^6\,3d\,n'l'$ series
limit at $19.884~\text{eV}$ \citep{Toresson1960} is clearly
discernable. The maximum of the DR rate coefficient at the
$2p^6\,3d\,n'l'$ series limit is about two orders of magnitude
smaller than that at the $2p^6\,3p\,n'l'$ series limit. The energy
range $20$--$69~\text{eV}$ is not shown because there is no
significant structure exceeding
$10^{-12}~\text{cm}^3\,\text{s}^{-1}$, but in principle it comprises
$2p^6\,nl\,n'l'$ resonances associated with $3s\to nl$ excitation
with $n\ge4$.
\begin{figure}[htb]
\includegraphics[width=\columnwidth]{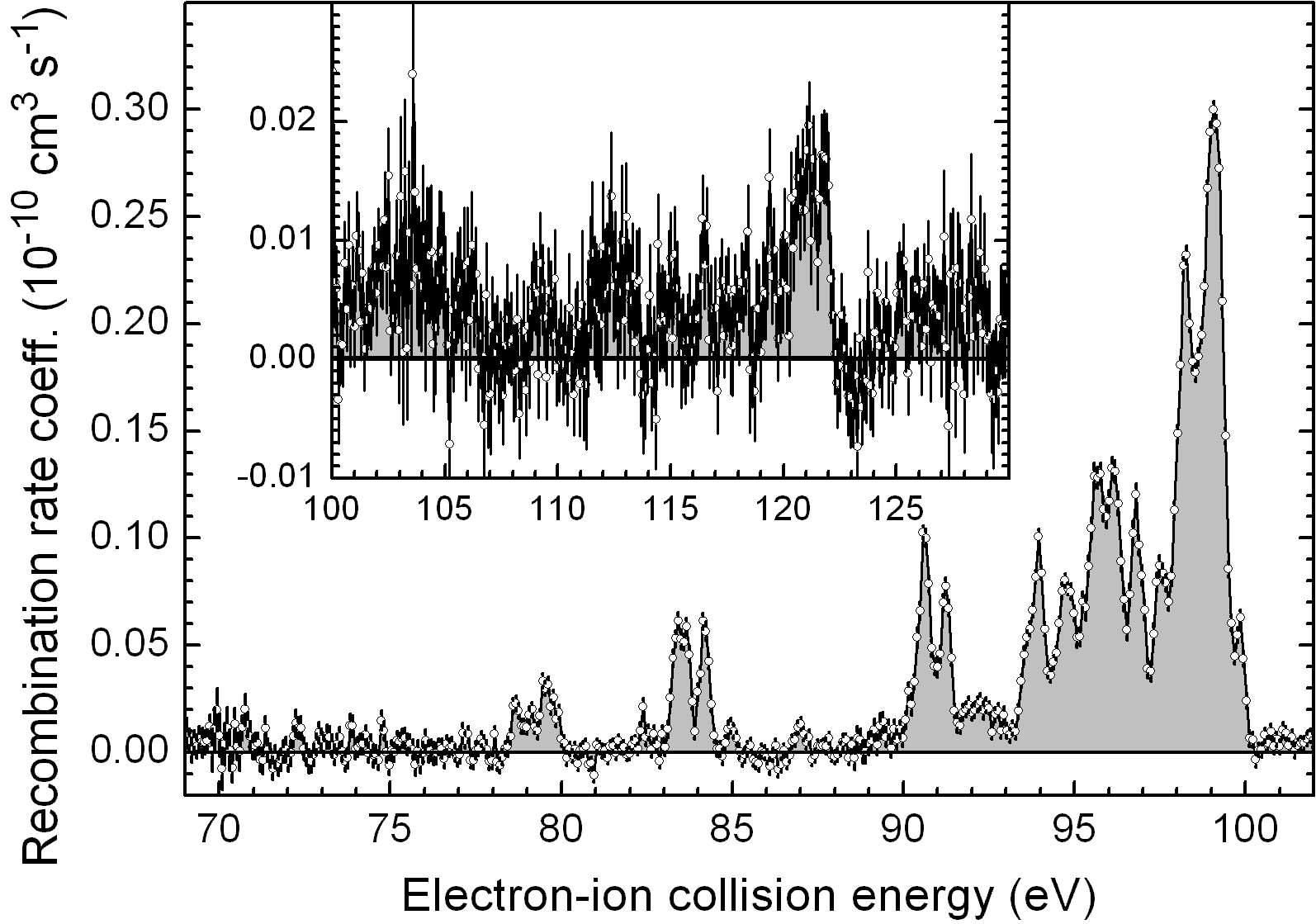}
\caption{Measured \ionsmall{Si}{4} to \ionsmall{Si}{3} merged-beams
recombination rate coefficient in the electron-ion collision-energy
region of resonances associated with $\Delta N\geq1$ $2s\to nl$ and
$2p\to nl$ ($n\geq 3$) inner shell core
excitations.\label{fig::DN1}}
\end{figure}

In Figure~\ref{fig::DN1} we show the measured merged-beams
recombination rate coefficient in the energy range
$69$--$130~\text{eV}$. This was measured using the thermionic
cathode. The resonances in this energy range are associated with
$\Delta N\geq1$ DR via $2p\to nl$ ($n\geq 3$) inner shell core
excitations. Due to the large number of excitations in the displayed
energy range (cf.\ Table~\ref{table::excenergy}) it is prohibitive to
assign individual resonances. The main contribution to the rate
coefficient is most probably due to $2p\to 3l$ core excitations.
Theoretical calculations suggest that DR by $2s$ excitations is
insignificant \citep{Altun2006a}. The energy range from $130$ up to
$186~\text{eV}$ was also scanned and found to fluctuate with
peak-to-peak variations up to
$\pm10^{-12}~\text{cm}^3\,\text{s}^{-1}$, not showing any significant
structure.

\begin{figure}[htb]
\includegraphics[width=\columnwidth]{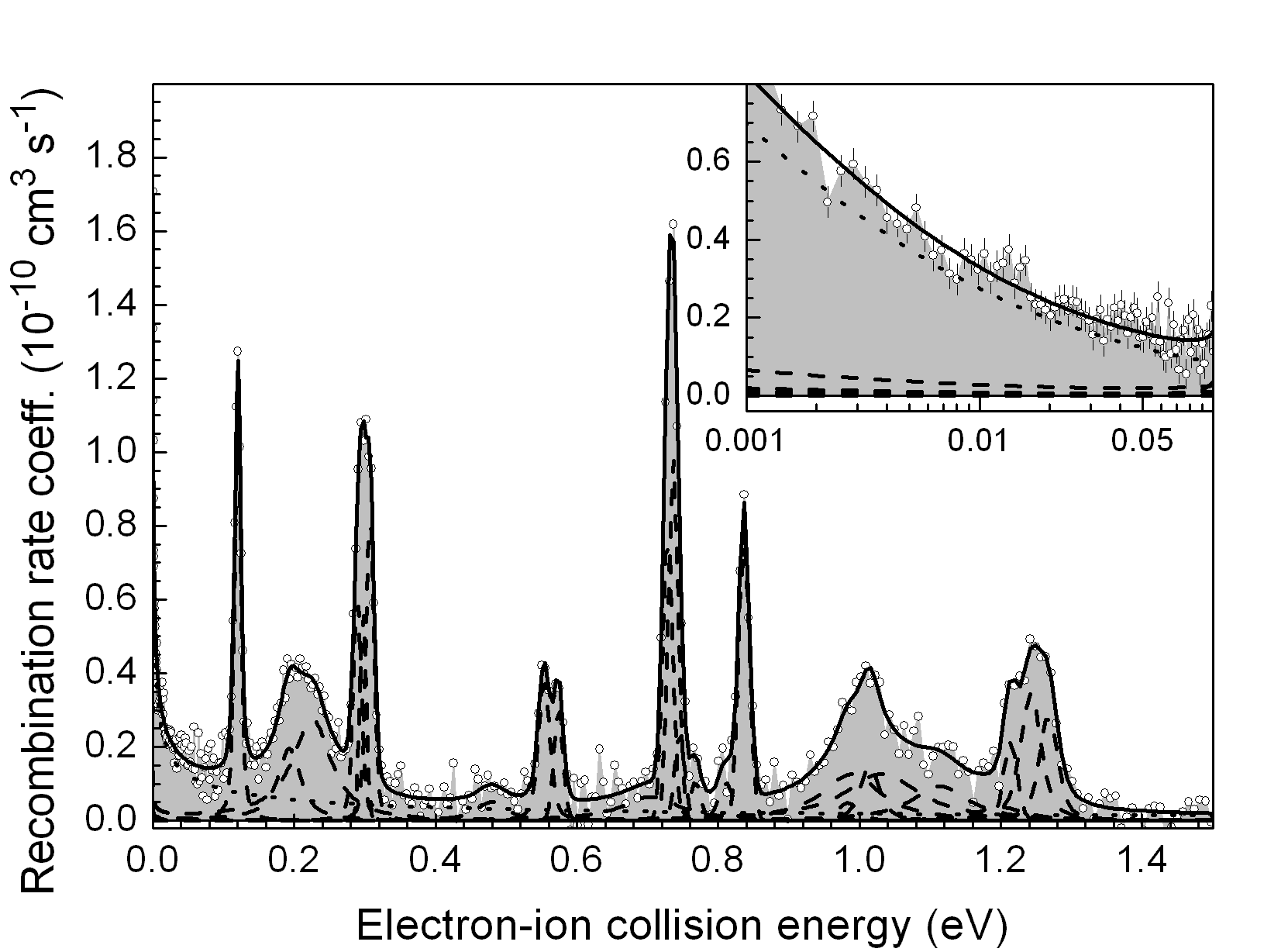}
\caption{Measured \ionsmall{Si}{4} to \ionsmall{Si}{3} merged-beams
rate coefficient at energies below $1.5~\text{eV}$ (circles and grey
shaded area). The solid curve is the sum of a fit comprising 25
fitted DR resonances (dashed curves cf.\
Sec.~\ref{subsec::plasma_derivation}) and the non-resonant rate
coefficient due to RR (dotted curve). The inset shows the same
curves, but in more detail in the energy range
$1$--$100~\text{meV}$.}\label{fig::Si3Lowe}
\end{figure}

The calculated RR rate coefficient as well as 25 DR
resonances fitted to to the measured DR spectrum in the energy region
below 1.5 eV  is shown in Figure~\ref{fig::Si3Lowe}. This fit is
independent of the theoretical predictions and is explained in detail
in Section~\ref{subsec::plasma_derivation}.
For the comparison with the theoretical calculation shown in the
Figures~\ref{fig::DN0theory}, \ref{fig::Si3DN0FritzscheFinal}, and
\ref{fig::DN1theory} the non-resonant RR contribution was subtracted
from the measured merged-beams recombination rate coefficient at all
energies. The merged-beams RR rate coefficient was derived by
convolving the RR cross section with the experimental electron
energy distribution. The RR cross section was calculated with a
hydrogenic formula \citep[Eq.~(12) of Ref.][]{Schippers2001c},
taking
into account field ionization of loosely bound high Rydberg
electrons inside the storage ring bending magnets
\citet{Schippers2001c}.

\subsection{Theoretical results}\label{subsec::theoresult}
To simulate the observed DR spectra at different energies of the
initially free electron, detailed computations have been carried out
for the $3s \to 3p\,n'l'$ $(n'=3,\ldots,6)$ and $3s \to 3d\,n'l'$
$(n'=3,\ldots,6)$ resonant electron capture and its subsequent
radiative stabilization. The first DR resonance ($3p\,4d\,^1D_2$) is
found at about $0.1~\text{eV}$, i.e., very close to zero energy of
the incident electron. The exact energies and strengths of resonances
close to the threshold are particularly important for the
determination of reliable plasma recombination rate coefficients at
low electron temperatures. Therefore, special care was taken with
regard to the doubly-excited magnesium-like states from the
$3l\,n'l'$ $(n' \ge 3)$ final configurations of the recombined ion.

For the $0$--$6~\text{eV}$ low-energy part of the DR spectrum, a
series of computations has been carried out. In the first approach,
we included all the levels within the $3l\,3l'$ and $3l\,4l'$
configurations of the recombined ion. Apart from these low-lying
levels with energies both below and above the threshold, we
incorporated in a further step also the levels of the $3l\,5l'$ and
$3l\,6l'$ configurations as well as later all the $nl\,n'l'$ levels
with $n \le 5$ and $n' \le 6$, respectively. To keep the number of
CSF manageable, levels with higher $n$ and $n'$ were not treated. As
known from previous computations \citep{Dong1999, Fritzsche2000b} for
the low-lying levels of multiply charged ions, such a systematic
enlargement of the wave function expansion (\ref{eq::ASFansatz})
typically improves the positions of the resonances significantly. For
magnesium-like ions, moreover, many of the $3l\,n'l'$ configurations
`overlap' with each other in energy and, hence, `new' resonances may
appear in the calculated low-energy part of the theoretical DR
spectrum, if the configuration space is increased. Using the single
and double excitations from above, we obtained an expansion of up to
1073 CSF for the intermediate and the final-state wave functions of
the \ion{Si}{3} ions.

For the $2p \to 3l\,n'l'$ part of the DR spectrum we could include
only levels of $2p^5\,3s\,3l\,3l'$ and $2p^5\,3s\,3l\,4l'$, since the
number of open shells involved is increased as compared to the $3s
\to 3l\,n'l'$ excitations. In the calculation the associated
resonances appeared in the energy region $69$--$94~\text{eV}$ of the
incident electrons. For these inner-shell excited spectra, further
contributions from the core polarization or core--core excitations
need to be omitted owing to the size requirements of the
corresponding wave function expansions. The incorporation of double
excitations from the $2s$ and $2p$ shells would result in expansions
of several hundred thousand CSF, i.e., a size which is unfeasible for
the computation of DR and autoionization properties.

\begin{figure}[htb]
\includegraphics[width=\columnwidth]{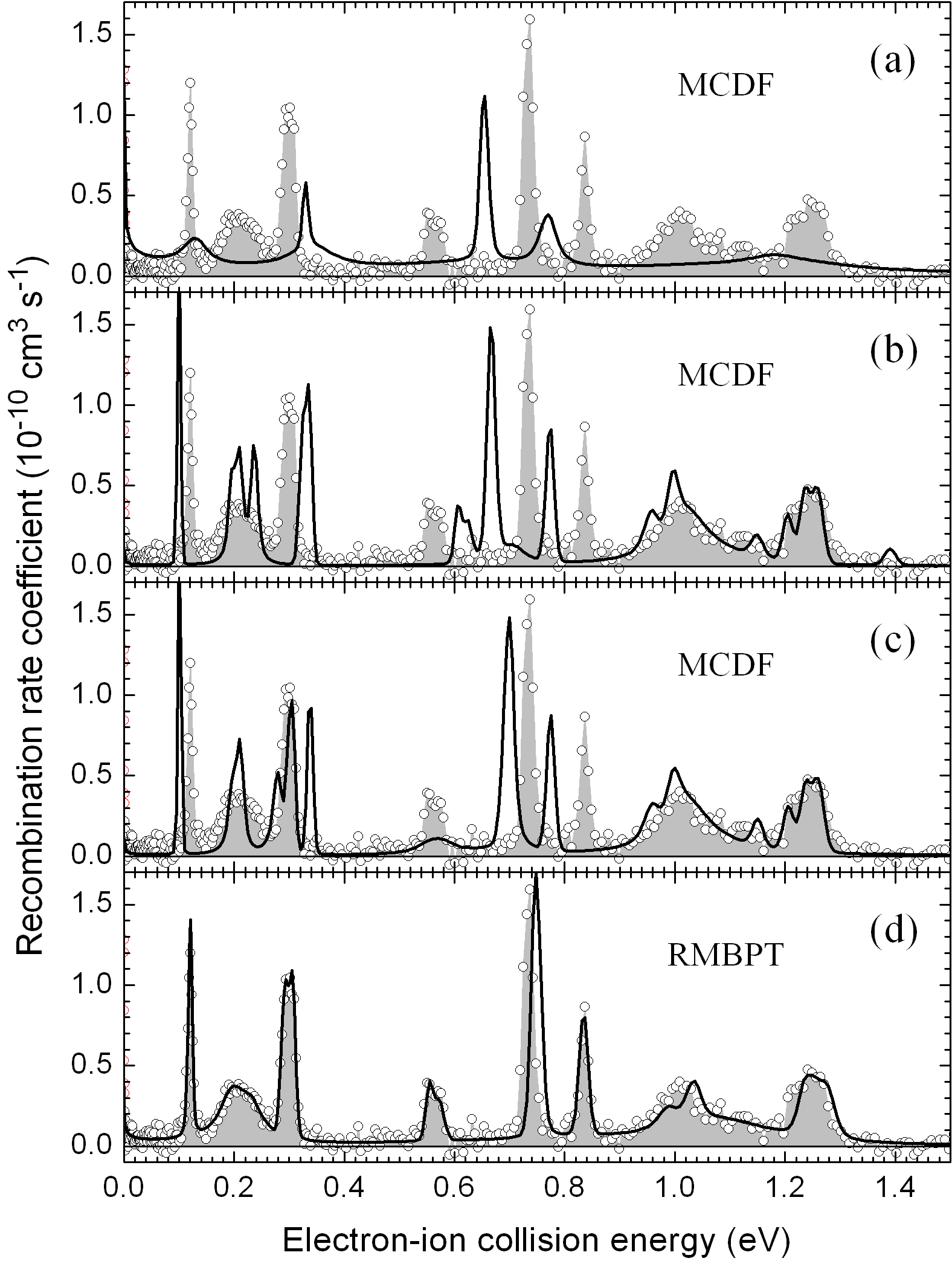}
\caption{Comparison of the experimental \ionsmall{Si}{4} DR spectrum
(open circles) in the energy region $0-1.5~\text{eV}$ with various
theoretical results (solid curves). Figures (a), (b), and (c) show
our multiconfiguration Dirac-Fock (MCDF) method results and (d)
published relativistic many-body perturbation theory (RMBPT)
calculations (Ref.~\citep{Orban2007a}). The nonresonant part of the
recombination rate coefficient due to radiative recombination (RR)
was subtracted from the measured curve (see text). In the MCDF
calculation three different approximations have been applied to the
enlarged CSF basis including all the fine-structure states from the
$nl\,n'l'~(n,n' = 3,4,5,6)$ configurations: (a) using the orbital
functions from the doubly-excited $3l\,n'l'\,^{2S+1}L_J$ levels for
the representation of both, the initial and final states; (b) with an
independent optimization of the sodium- and magnesium-like states but
without including the exchange interaction between the initially free
electron and the bound-state electrons; (c) the same as in (b) but by
incorporating the exchange interaction with regard to the bound-state
density.} \label{fig::DN0theory}
\end{figure}

Figure~\ref{fig::DN0theory} displays the experimental \ion{Si}{4} DR
spectrum in the energy region of $0$--$1.5~\text{eV}$, with the
nonresonant `background' subtracted from the experimental results. In
this figure, the observed spectrum is compared with theoretical
results from different approximations. In all of these calculations,
our `best' wave function expansion has been applied, including the
single and double excitations as discussed above. The computations
differ however in the set of the one-electron orbital functions used
for the representation of the initial sodium-like ions. Figure~
\ref{fig::DN0theory}(a), for instance, shows the spectrum in which
both the initial and final states of the recombined ion were
described by a \emph{common} set of orbitals, neglecting the
rearrangement of the electron density in the course of the
dielectronic capture (or decay) of the ions. Apparently, quite a
strong effect arises from this rearrangement of the electron density
as seen from Figure~\ref{fig::DN0theory}(b) and
\ref{fig::DN0theory}(c), for which two independent sets of orbital
functions were utilized in the representation of the initial and the
recombined ion states. Figures~ \ref{fig::DN0theory}(b) and
\ref{fig::DN0theory}(c) only differ in the treatment of the exchange
interaction for the incoming electron. While
Figure~\ref{fig::DN0theory}(b) shows the simulation for a static
potential due to the charge distribution of the initial ion (`no
exchange'), \ref{fig::DN0theory}(c) incorporates the exchange
interaction of the incident electron with regard to the bound-state
density. Therefore, Figure~\ref{fig::DN0theory}(c) represents our
best approximation within the MCDF approach. Despite the fact that
the orbitals are not quite orthogonal in the computation of the
two-particle matrix elements, the Auger amplitudes
[Eq.~(\ref{eq::Aa-rate})] were evaluated by using the techniques of
Racah's algebra, i.e., for assuming orthogonality for all
\emph{inactive} electrons in these transition amplitudes
\cite{Fritzsche2007a}.

Figure~\ref{fig::DN0theory}(d) displays the comparison of the present
experimental data with the RMBPT result of Orban et al.\
\cite{Orban2007a} convolved with the electron energy distribution of
the TSR photocathode electron beam. Compared to their experiment,
where electron beam temperatures
$k_\text{B}T_\parallel=0.25~\text{meV}$ and
$k_\text{B}T_\perp=10~\text{meV}$ were found, the energy resolution
is higher in the present experiment.
($k_\text{B}T_\parallel=0.035~\text{meV}$,
$k_\text{B}T_\perp=0.9~\text{meV}$, see
Section~\ref{sec::experiment}). Our high-resolution experiment
provides a more stringent test of the RMPBT calculation which
represents the measurement almost perfectly but cannot easily be
extended to energies beyond $1.4~\text{eV}$ where an increasing
number of resonances can contribute to the DR spectrum.
%
\begin{table}[htb]
\caption{\label{table::calc} Comparison of the present
multiconfiguration Dirac-Fock (MCDF) resonance parameters with the
results of the relativistic many-body perturbation theory (RMBPT,
Ref.\ \citep{Orban2007a}) for all $3p\,4l$ and $3p\,5s$ DR resonances
associated with $3s\to 3p$ core excitations. The listed quantities
are resonance energies $E_\text{res}$ and strengths $S$
[Eq.~(\ref{eq::resonance-strength})]. The dominant LS-terms of the
resonance states are given in the first column along with their
weights in the representation of the wave functions. The weights are
derived from the MCDF calculation. The states are listed in the order
of increasing MCDF resonance energies. RMBPT resonance energies which
appear in a different order \citep{Orban2007a} are marked by an
asterisk in the third column.}
\begin{ruledtabular}
\begin{tabular}{l dd dd}
 \multicolumn{1}{c}{dominating}
  & \multicolumn{2}{c}{$E_\text{res}$ (eV)}
   & \multicolumn{2}{c}{$S$ ($10^{-20}$ eV cm$^2$)} \\
 \cline{2-3}\cline{4-5}
 \vspace{-2ex}\\
 \multicolumn{1}{c}{LS term}
  & \multicolumn{1}{c}{MCDF}   &  \multicolumn{1}{c}{RMBPT}
   & \multicolumn{1}{c}{MCDF}   &  \multicolumn{1}{c}{RMBPT}  \\
 \hline
 \vspace{-2ex}\\
 $3p\,4d\,\,^1D_2\;\hspace{\fill}(96\%)$ & 0.102& 0.121& 8.33& 6.88\\
 $3p\,4d\,\,^3F_2\;\hspace{\fill}(94\%)$ & 0.198& 0.191& 3.49& 3.77\\
 $3p\,4d\,\,^3F_3\;\hspace{\fill}(97\%)$ & 0.212& 0.204& 4.55& 4.95\\
 $3p\,4d\,\,^3F_4\;\hspace{\fill}(99\%)$ & 0.280& 0.233& 4.46& 5.58\\
 $3p\,4d\,\,^3D_1\;\hspace{\fill}(95\%)$ & 0.299& 0.288& 1.83& 1.78\\
 $3p\,4d\,\,^3D_2\;\hspace{\fill}(94\%)$ & 0.307& 0.296& 2.96& 2.96\\
 $3p\,4d\,\,^3D_3\;\hspace{\fill}(94\%)$ & 0.339& 0.307& 3.76& 4.01\\
 $3p\,4d\,\,^3P_2\;\hspace{\fill}(95\%)$ & 0.557& 0.556& 1.57& 1.42\\
 $3p\,4d\,\,^3P_1\;\hspace{\fill}(95\%)$ & 0.583& 0.572& 0.91& 0.83\\
 $3p\,4d\,\,^3P_0\;\hspace{\fill}(95\%)$ & 0.633& 0.581& 0.28& 0.27\\
 $3p\,4f\,\,^3F_2\;\hspace{\fill}(93\%)$ & 0.691& 0.744& 1.38& 1.72\\
 $3p\,4f\,\,^3F_4\;\hspace{\fill}(77\%)$ & 0.702& 0.754& 3.10& 3.20\\
 $3p\,4f\,\,^3F_3\;\hspace{\fill}(92\%)$ & 0.702&*0.748& 2.65& 2.50\\
 $3p\,4f\,\,^1G_4\;\hspace{\fill}(68\%)$ & 0.709& 0.796& 1.70& 1.46\\
 $3p\,4f\,\,^1F_3\;\hspace{\fill}(99\%)$ & 0.775& 0.835& 3.14& 2.96\\
 $3p\,5s\,\,^3P_0\;\hspace{\fill}(73\%)$ & 0.945& 0.976& 0.25& 0.24\\
 $3p\,5s\,\,^3P_1\;\hspace{\fill}(72\%)$ & 0.960& 0.991& 0.74& 0.71\\
 $3p\,4f\,\,^3G_3\;\hspace{\fill}(98\%)$ & 0.995& 1.048& 2.33& 2.48\\
 $3p\,5s\,\,^3P_2\;\hspace{\fill}(74\%)$ & 0.999&*1.031& 1.20& 1.14\\
 $3p\,4f\,\,^3G_4\;\hspace{\fill}(98\%)$ & 1.013& 1.067& 2.94& 3.13\\
 $3p\,4f\,\,^3G_5\;\hspace{\fill}(98\%)$ & 1.034& 1.088& 3.54& 3.75\\
 $3p\,4d\,\,^1F_3\;\hspace{\fill}(79\%)$ & 1.087&*0.654& 1.58& 2.23\\
 $3p\,5s\,\,^1P_1\;\hspace{\fill}(73\%)$ & 1.151& 1.133& 0.84& 0.78\\
 $3p\,4f\,\,^3D_2\;\hspace{\fill}(85\%)$ & 1.207& 1.276& 1.10& 1.01\\
 $3p\,4f\,\,^3D_3\;\hspace{\fill}(97\%)$ & 1.238&*1.254& 1.66& 1.56\\
 $3p\,4f\,\,^1D_2\;\hspace{\fill}(90\%)$ & 1.258&*1.235& 1.16& 1.13\\
 $3p\,4f\,\,^3D_1\;\hspace{\fill}(97\%)$ & 1.266& 1.282& 0.70& 0.65\\
 $3p\,4d\,\,^1P_1\;\hspace{\fill}(73\%)$ & 1.393&*1.040& 0.46& 0.54\\
\end{tabular}
\end{ruledtabular}
\end{table}

Table~\ref{table::calc} displays the assignment and position of the
28 lowest resonances in the energy region $E_\text{res} \lesssim
1.5~\text{eV}$ calculated with the MCDF method as well as the
corresponding RMBPT results of \citet{Orban2007a}. In addition to the
energies of these resonances, this table also includes the weights of
the dominant $LS$ terms as well as the resonance strengths. The
weights of the leading LS terms have been obtained by a unitary
transformation of the wave functions from the $jj$-coupled into a
$LS$-coupled basis \cite{Gaigalas2004}. While the lowest 11
resonances appear rather pure in $LS$ coupling ($\gtrsim\, 92$~\%{}),
some larger admixtures are found for a few of the higher-lying
resonances. As mentioned above, all data in this table correspond to
our best representation of the resonances and by including the
effects of the rearrangement of the electron density and the exchange
interaction (cf.\ Fig.~\ref{fig::DN0theory}c). The lowest resonances
in the $3s \to 3p\,n'l'$ part of the DR spectrum belong to the
$3p\,4d\,^1D_2^o$ level, followed by the fine-structure levels of the
$3p\,4d\,^3F^o$ term. Apparently, all levels from the $3p^2$ and
$3p\,3d$ configurations are below the DR threshold.

At energies below $0.9~\text{eV}$  the RMBPT results of
\citet{Orban2007a} are in excellent agreement with the present
measurements [Fig.\ \ref{fig::DN0theory}(d)]. At higher energies,
where the MCDF results reproduce the experimental findings slightly
better than at lower energies, the RMBPT resonance positions are at
somewhat too high energies as can also be seen in the comparison with
the experimental data of \citet{Orban2007a}. In our work, this slight
discrepancy is more pronounced because of the increased experimental
resolution and because of reduced statistical uncertainties in the
present experiment. With a few exceptions, MCDF and RMBPT resonance
energies agree with one another to within $\sim 50~\text{meV}$, often
even to within $\sim 20~\text{meV}$.

\begin{figure}[htb]
\includegraphics[width=\columnwidth]{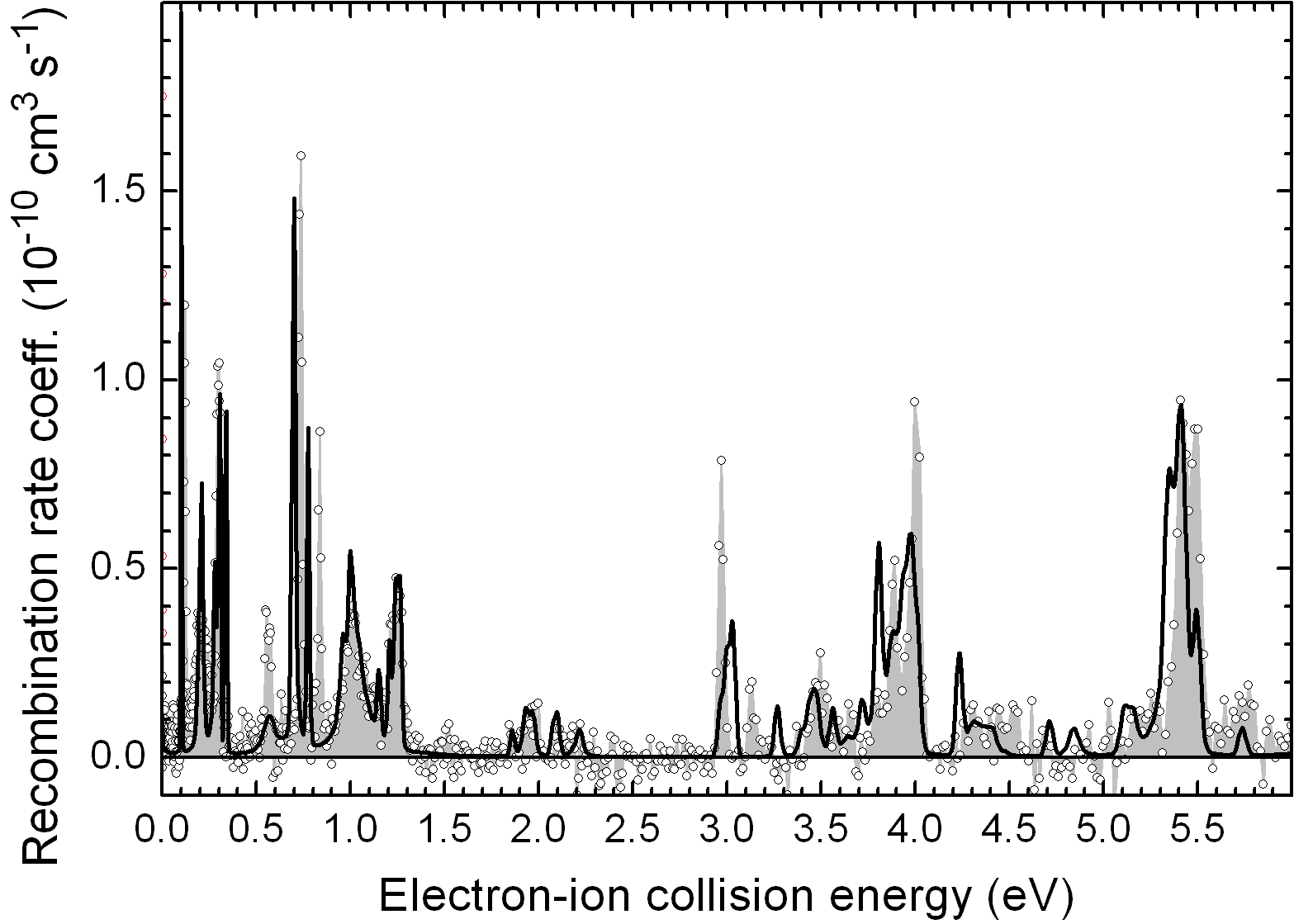}
\caption{Comparison of the experimental \ionsmall{Si}{4} DR spectrum
(open circles) in the energy region $0-6~\text{eV}$ with the final
result of the MCDF calculation (solid line). This energy range
includes $2p^6\,3p\,n'l'$ ($4\leq n'\leq6$) and $2p^6\,3d^2$
resonances. The nonresonant part of the recombination rate
coefficient due to RR was subtracted from
the measured curve (see text)}\label{fig::Si3DN0FritzscheFinal}
\end{figure}

\begin{figure}[htb]
\includegraphics[width=\columnwidth]{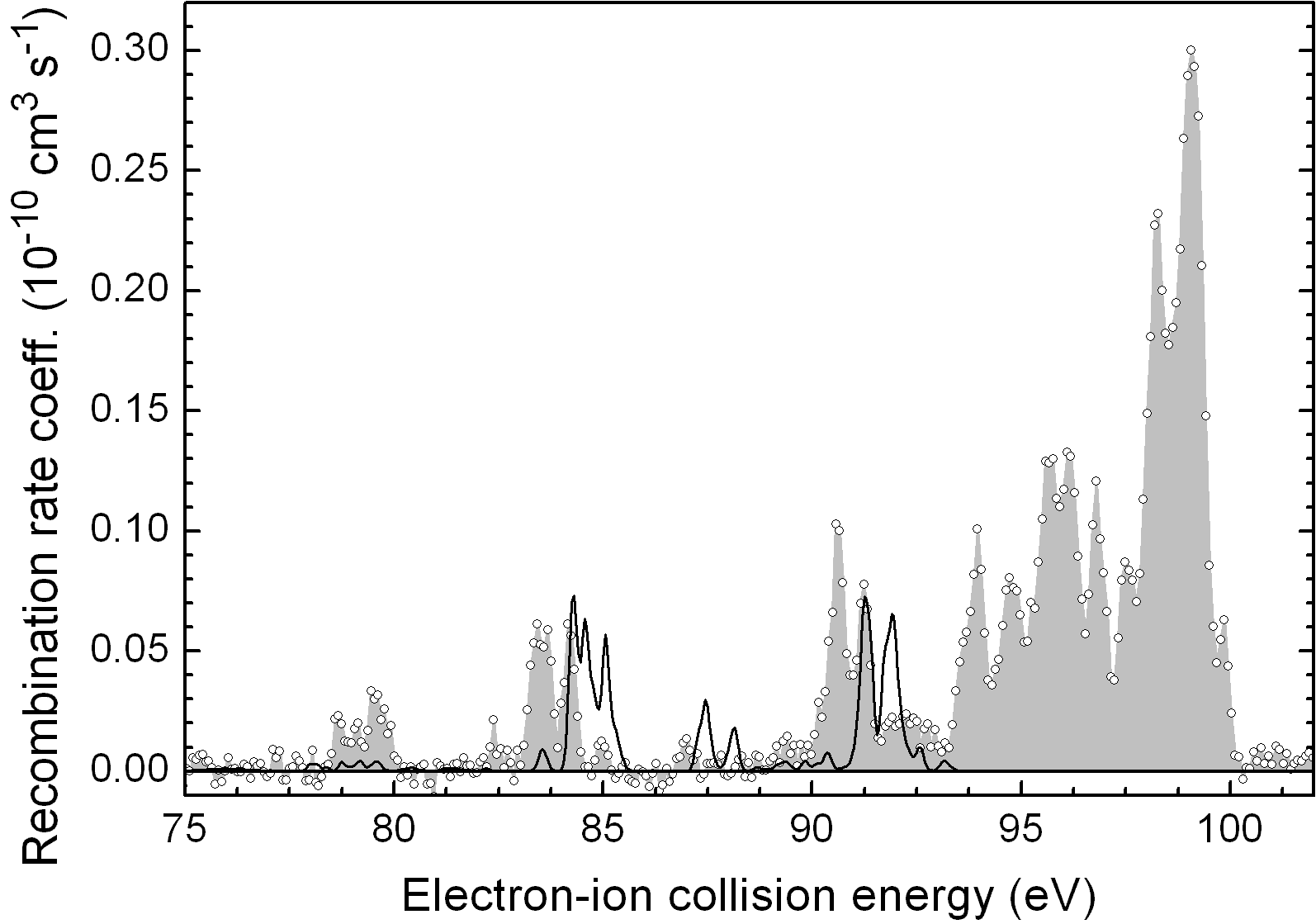}
\caption{Comparison of experimental \ionsmall{Si}{4} DR spectrum
(open circles) in the energy region of the $2p^5\,3s\,3l\,n'l'$
resonances with the results of our MCDF calculation (solid line).
The nonresonant part of the recombination rate coefficient due to RR
was subtracted from the
measured curve (see text)}\label{fig::DN1theory}
\end{figure}

Figures~\ref{fig::Si3DN0FritzscheFinal} and \ref{fig::DN1theory}
display our theoretical MCDF DR spectrum compared with experiment for
$\Delta N=0$ and $\Delta N=1$ DR. As described above, all $nl\,n'l'$
configurations with $3\le n,n' \le 6$ have been taken into account
for $\Delta N=0$ DR. The incorporation of further configurations with
even higher principal quantum numbers $n'$ has no effect upon the
low-lying resonances for energies $E_\text{res}\lesssim 6~\text{eV}$
above the threshold. Although the basic features are well described
in these spectra, some deviations in the positions and strengths of
the peaks remain which we attribute to neglected correlation and
many-electron effects in the system.

In addition to calculations of the low-lying resonances associated
with $3s \to 3p$ and $3s \to 3d$ core excitations, computations have
been carried out also for all $2p^5\,3s\,3l\,3l'$ and
$2p^5\,3s\,3l\,4l'$ resonances which are found to occur starting at
$69~\text{eV}$. For this high-energy part of the spectrum, we expect
only a rough agreement between our computations and experiment as the
electronic structure of the intermediate resonances now includes
\emph{four} open shells which do not allow any systematic enlargement
of the configuration basis. For this part of the spectrum, therefore,
the computations have been restricted to allow only one electron in
the $n=4$ shell. As seen in Figure~ \ref{fig::DN1theory}, only some
of the resonances at $84$ and $91~\text{eV}$ are reproduced by our
simulations but they are shifted upwards in energy by about
$1~\text{eV}$.

Apart from the radiative stabilization of the intermediate resonances
by E1 electric-dipole decay, we explored also the effects of higher
multipoles (M1, E2, M2) in the coupling of the radiation field. These
`multipoles' are well suppressed for neutral systems by at least 5
orders of magnitude but become important for highly-charged ions. For
the initially triply ionized \ion{Si}{4} ions, these contributions
are still negligible at the present level of accuracy for the
simulation of the DR spectra in Figures~
\ref{fig::DN0theory}--\ref{fig::DN1theory}.

\section{Plasma rate-coefficients}\label{subsec::plasma}
DR rate coefficients for a Maxwellian plasma can be derived from the
experimental merged-beams recombination rate coefficient and the
theoretical cross section. This is done in the following sections. Of
particular interest for astrophysical model calculations are the
plasma DR rate coefficients in the electron temperature ranges where
\ion{Si}{4} is formed in astrophysical plasmas. The approximate
temperature range where \ion{Si}{4} forms in photoionized and
collisionally ionized plasmas can be obtained from the work of
\citet{Kallman2001} and \citet{Bryans2006}, respectively. For
photoionized plasmas, \citet{Kallman2001} find that the fractional
\ion{Si}{4} abundance peaks at a temperature of $1~\text{eV}$. The
`photoionized zone' may be defined as the temperature range where the
fractional abundance of a given ion exceeds 10\% of its peak value.
For \ion{Si}{4} this corresponds to a temperature range of
$0.8$--$1.4~\text{eV}$. Using the same criterion and the fractional
abundances of \citet{Bryans2006}, for coronal equilibrium the
\ion{Si}{4} `collisionally ionized zone' is estimated to extend over
a temperature range of $4$--$10~\text{eV}$. It should be kept in mind
that these temperature ranges are only indicative. They depend, in
part, on the accuracy of the underlying atomic data base.

\subsection{Derivation of the plasma DR rate
coefficients}\label{subsec::plasma_derivation}
The DR rate coefficient in a Maxwellian plasma is derived by
convolving the DR cross section $\sigma_\text{DR}$ with an isotropic
Maxwell-Boltzmann electron energy distribution as detailed by
\citet{Schippers2001c, Schippers2004c}. To derive a meaningful
plasma DR rate coefficient from a total merged-beams rate
coefficient there are some issues that require special
consideration.

Interference between DR and RR is typically unimportant
\citep{Pindzola1992a}. Hence here we subtract, the non-resonant RR
contribution from the measured merged-beams recombination rate
coefficient. The applied merged-beams RR rate coefficient was the
same as used for the correction of the recombination rate coefficient
[Eq.~(\ref{eq::alpha})] at the reference point as well as that which
was subtracted from the experimental merged-beams recombination rate
coefficient for comparison with theoretical results (cf.\
Sec.~\ref{subsec::expresults}).

When the electron-ion collision energy $E$ is larger than the
experimental energy spread $\Delta E$, one can use
$\alpha_\text{DR}/(2 E/m_\text{e})^{1/2}$ instead of
$\sigma_\text{DR}$ for the convolution. When $E\lesssim\Delta E$, the
energy spread influences the outcome of the convolution of the cross
section. In order to account for this effect, the low energy DR cross
section was extracted by fitting $25$ DR resonance line-shapes to the
measured DR spectrum in the energy region below $1.5~\text{eV}$
(Fig.\ \ref{fig::Si3Lowe}), independent of the theoretical
predictions above (see Ref. \citep{Schippers2004c} for a more
detailed description of the method).

\begin{figure}[htb]
\includegraphics[width=\columnwidth]{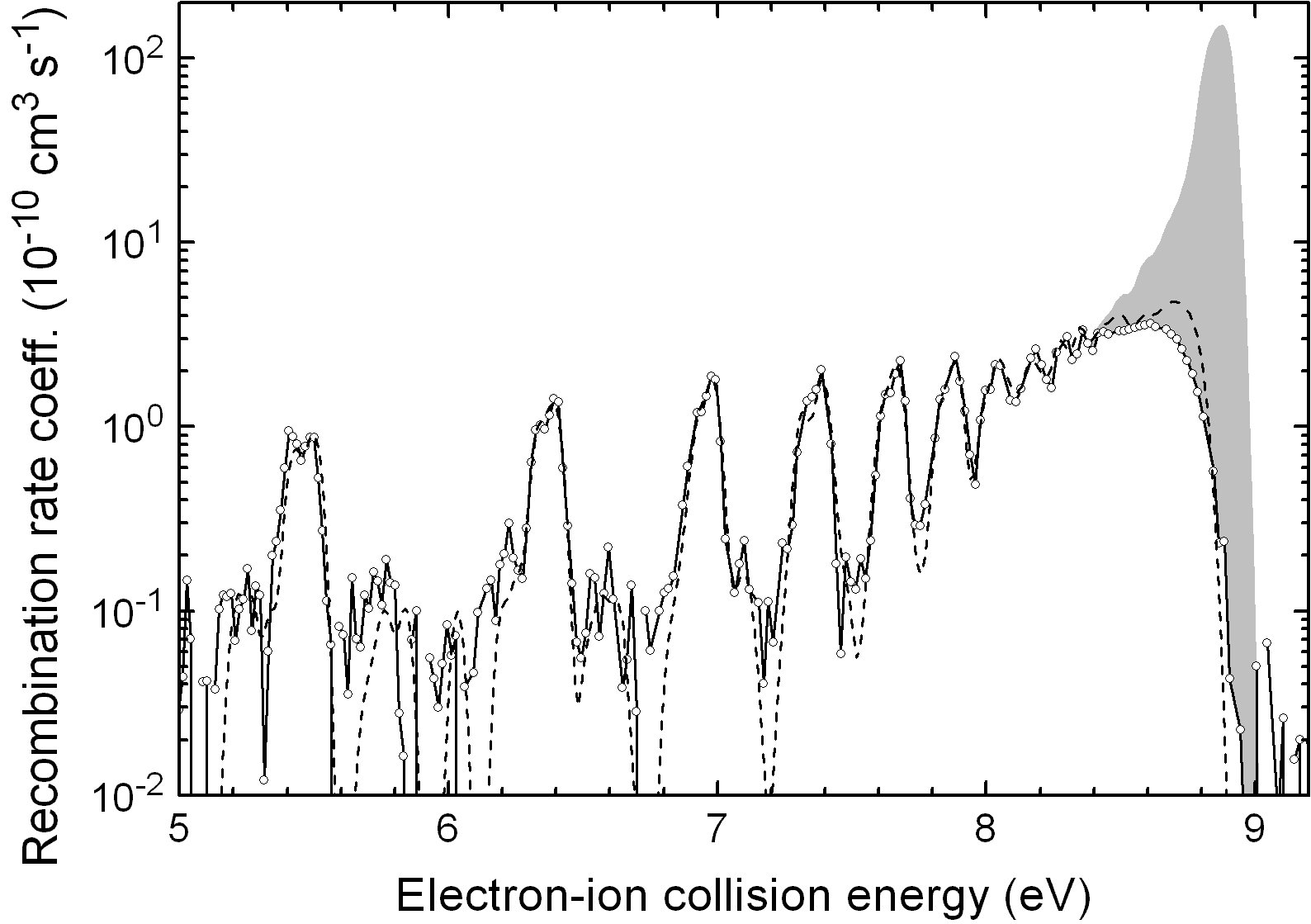}
\caption{Comparison between the experimental merged-beams DR rate
coefficient (open circles with solid line) and the AUTOSTRUCTURE
calculation. The AUTOSTRUCTURE calculation was multiplied by a factor
of 1.13 (see text). The dashed line is the theoretical result with
account for the experimental field ionization of high-$n$ Rydberg
states using the field ionization model of \citep{Schippers2001c}.
The shaded area highlights the unmeasured purely calculated part of
the composite DR rate coefficient.}\label{fig::Si3vsautos1000}
\end{figure}
Field ionization of the loosely bound high Rydberg electron in the
recombined ions can result from the motional electric fields that the
ions experience inside the storage ring bending magnets
\citep{Schippers2001c}. The ion beam on its way to the detector
passes the strongest electrical field in the dipole bending magnet in
front of the detector. From this electric field the critical quantum
number for field ionization is $n_\text{crit}=17$, i.e., in the
present experiment only RR and DR involving capture into Rydberg
levels with quantum numbers less than $17$ contribute fully to the
measured merged-beams recombination rate coefficient. Due to
radiative decay of higher Rydberg states on the way from the
interaction section to the detector, the field ionization cutoff is
not sharp but somehow smeared out to $n\geq17$. Similar to the
approach of \citet{Schippers2001c, Schippers2004c} the missing DR
resonance strength up to $n_\text{max}=1000$ was estimated from a
theoretical calculation using the AUTOSTRUCTURE code
\citep{Badnell1986}. $n_\text{max}=1000$ is an arbitrary upper limit
beyond which no significant contribution to the total DR cross
section is expected. Although the AUTOSTRUCTURE code does not
reproduce the resonance structure below $5~\text{eV}$ (cf. Fig.~1 of
\citep{Orban2006a}) as accurately as our present MCDF calculations,
it reproduces the more regular structures of Rydberg resonances
between $5~\text{eV}$ and $8.4~\text{eV}$ when the calculated rate
coefficient is multiplied by a constant factor of $1.13$. The
unmeasured DR contribution due to $n\geq17$ exceeds the measured
contribution by more than one order of magnitude. This is shown in
Figure~\ref{fig::Si3vsautos1000}. The DR contribution for $\Delta
N=0$ $3s\to 3p$ DR from $n=17$--$1000$ was added to the measured
spectrum by adding the difference between the measured rate
coefficient and the adjusted AUTOSTRUCTURE result in the energy range
$8.39~\text{eV}-9.01~\text{eV}$ (grey shaded area in
Figure~\ref{fig::Si3vsautos1000}).

The $3s\to 3d$ $\Delta N=0$ series with its limit at about
$20~\text{eV}$ was not corrected for field ionization losses of high
Rydberg states because its contribution is negligible compared to the
$3s\to 3p$ $\Delta N=0$ series limit. As the strengths of resonances
contributed by $\Delta N\geq1$ DR fall much faster with increasing
$n$ than the strength of $\Delta N=0$ DR resonances, and because of
the much smaller contribution of $\Delta N\geq1$ DR, the field
ionized contribution for $\Delta N\geq1$ DR with $n\geq17$ was also
not corrected for field ionization losses, either.

The non-measured contribution to the plasma DR rate coefficient
ranges in the photoionized zone from only $1\%$ at $0.8~\text{eV}$ to
$36\%$ at $1.4~\text{eV}$. It has its maximum ($83\%$) at a plasma
electron temperature of $k_\text{B} T_\text{e}=7.9~\text{eV}$. In the
collisionally ionized zone the contribution ranges from $79\%$ at
$4~\text{eV}$ to $83\%$ at $10~\text{eV}$. The contribution falls off
slightly towards higher temperatures and is still $72\%$ at
$1000~\text{eV}$. The resulting plasma DR rate coefficient with and
without the correction of field ionization losses is shown in
Figure~\ref{fig::plasma}.

For convenient use in astrophysical modeling codes the \ion{Si}{4} to
\ion{Si}{3} plasma DR rate coefficient $\alpha_\text{p}^\text{(DR)}$
was fitted using
\begin{equation}
\alpha_\text{p}^\text{(DR)}(T_\text{e})=(T_\text{e})^{-3/2}\sum_{i=1}^{9}
c_i \exp(-E_i/k_\text{B} T_\text{e}). \label{eq::PDRfit}
\end{equation}
The fitting parameters $c_i$ and $E_i$ are given in
Table~\ref{table::PDRfit}. The fit deviates by less than $1\%$ from
the experimentally-derived result in the temperature range
$0.01~\text{eV}-10000~\text{eV}$.
\begin{table}[htb]
\caption{\label{table::PDRfit}Parameters for the fit of
Equation~(\ref{eq::PDRfit}) to the experimental plasma DR rate
coefficient. Numbers in square brackets denote powers of $10$.}
\begin{ruledtabular}
\begin{tabular}{cdd}
 $i$ &
\multicolumn{1}{r}{$c_i~(\text{cm}^3\,\text{s}^{-1}\,\text{K}^{3/2})$}
& \multicolumn{1}{r}{ $E_i~(\text{eV})$}\\
 \hline
 \vspace{-2ex}\\
 1 & 2.13[-8] & 1.02[-2]\\
 2 & 6.12[-8] & 5.00[-2]\\
 3 & 1.10[-6] & 1.24[-1]\\
 4 & 3.65[-6] & 2.44[-1]\\
 5 & 1.45[-5] & 6.90[-1]\\
 6 & 1.78[-5] & 1.53[+0]\\
 7 & 3.05[-4] & 5.43[+0]\\
 8 & 9.50[-3] & 8.81[+0]\\
 9 & 1.89[-3] & 8.05[+1]\\
\end{tabular}
\end{ruledtabular}
\end{table}

\subsection{Comparison with present
theory}\label{subsec::presentcomparison}
In Figure~\ref{fig::plasma} we compare the plasma DR rate coefficient
derived from the MCDF calculation in the electron-ion collision
energy range $0$--$6~\text{eV}$ and a plasma DR rate coefficient
derived from DR merged-beams resonances measured in the same
electron-ion collision energy range. We find good agreement in the
comparison of these two plasma DR rate coefficients. The plasma DR
rate coefficient generated by DR resonances, calculated by means of
the MCDF method for electron-ion collision energies below
$6~\text{eV}$ is somewhat lower than the rate coefficient generated
from the experimentally-derived resonances in the same energy range.
Between $k_\text{B} T_\text{e}=0.01~\text{eV}$ and $k_\text{B}
T_\text{e}=0.04~\text{eV}$ the agreement is better than $12\%$. Above
a plasma electron temperature of $k_\text{B} T_\text{e}
=0.04~\text{eV}$ the agreement is even better than $7\%$.

\subsection{Comparison with previous
results}\label{subsec::previouscomparison}
\begin{figure}[htb]
\includegraphics[width=\columnwidth]{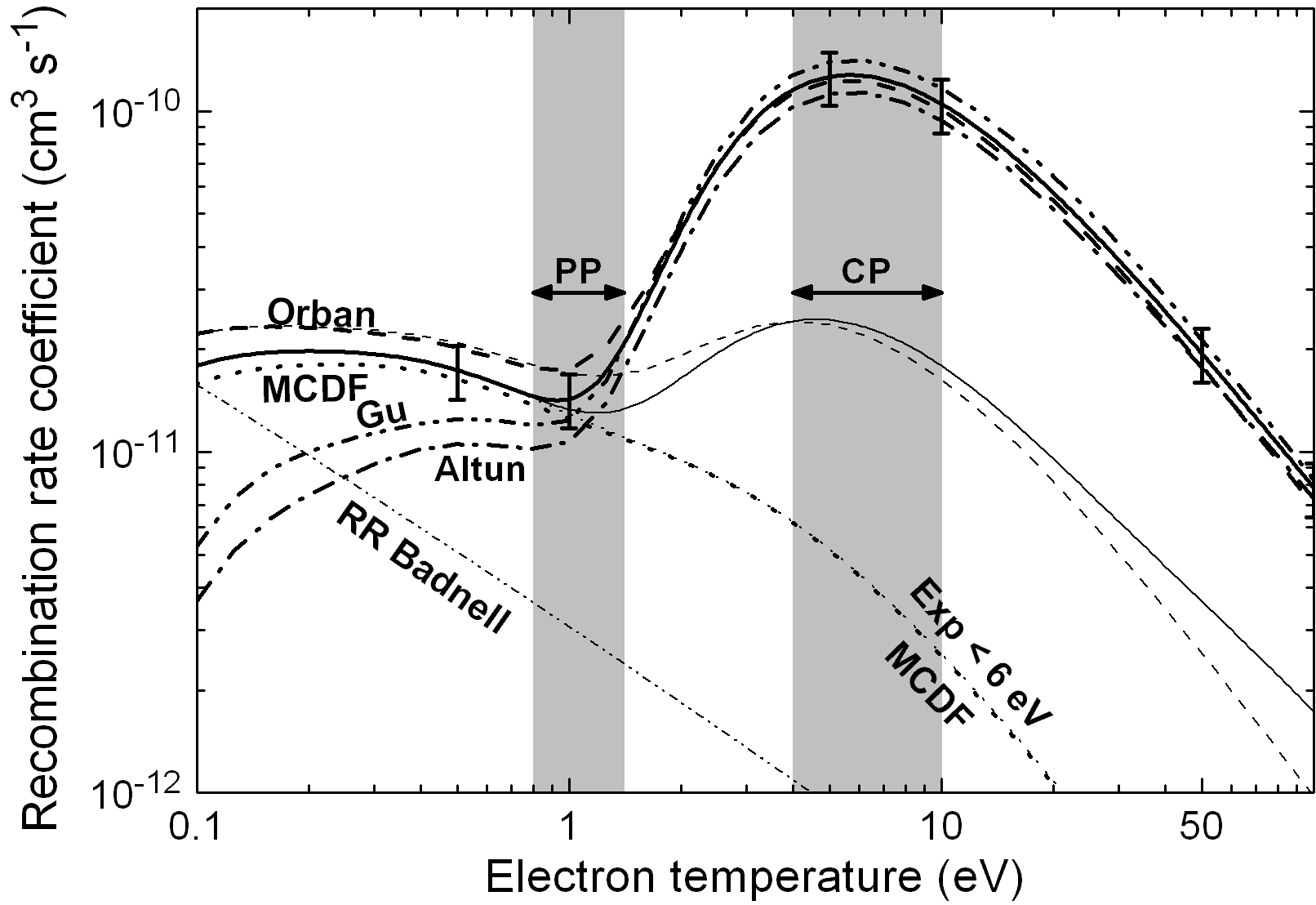}
\caption{Experimentally derived \ion{Si}{4} to \ion{Si}{3} DR rate
coefficient in a plasma (thick solid line) comprising $\Delta N=0$ DR
(Figs.~\ref{fig::DN0sp} and \ref{fig::DN0sd}), $\Delta N=1$ and $2$
DR (Fig.~\ref{fig::DN1}), and the theoretical estimate for the
unmeasured contributions of states with $n\geq17$ for $\Delta N=0$ DR
(Fig.~\ref{fig::Si3vsautos1000}). The error bars denote the $\pm 18\%
$ ($1\sigma$) experimental uncertainty in the absolute rate
coefficient. The experimental results without DR extrapolation is
shown by the thin solid line. Also shown are recent theoretical
calculations of the DR rate coefficient by \citet{Gu2004a} (thick
dash-dot-dotted line, labeled Gu) and \citet{Altun2006a} (thick
dash-dotted line, labeled Altun), and the recent experimental DR rate
coefficient by \citet{Orban2006a} (thick dashed line, labeled Orban).
The contribution from the experimentally measured DR resonances
between $0$ and $6~\text{eV}$ is shown as the thin dotted line
(labeled Exp~$<6~\text{eV}$) while the contribution of the MCDF
calculation in the same energy range is shown as thick dotted line
(labeled MCDF). A recent calculation of the plasma RR by
\citet{Badnell2006d} is shown as the thin dash-dot-dotted line
labeled RR Badnell. The temperature ranges where \ion{Si}{4} is
expected to peak in abundance in photoionized plasmas (labeled PP)
and collisionally ionized plasmas (labeled CP) are highlighted.}
\label{fig::plasma}
\end{figure}
In Figure~\ref{fig::plasma} we also compare our
experimentally-derived DR rate coefficient with the experimental
result of \citet{Orban2006a} and with recent theoretical results of
\citet{Gu2004a} and \citet{Altun2006a}.

In the collisionally ionized zone the rate coefficient of
\citet{Gu2004a} overestimates our experimentally-derived plasma DR
rate coefficient by about $10\%$. In contrast, the plasma DR rate
coefficient of \citet{Altun2006a} underestimates the
experimentally-derived plasma DR rate coefficient in the
collisionally ionized zone by about $10\%$. These deviations are
within the experimental uncertainty. The difference between the two
codes of about $20\%$ also gives an idea as to the uncertainty of
these state-of-the-art DR calculations.

In the photoionized zone the experimentally-derived plasma DR rate
coefficient is decisively determined (between $93\%$ at
$0.8~\text{eV}$ and $41\%$ at $1.4~\text{eV}$) by the $3s+e^-\to
3p\,4l'$ resonances which occur at electron-ion collision energies
below $1.5~\text{eV}$. The calculations of \citet{Gu2004a} and
\citet{Altun2006a} comprise DR for all $\Delta N=0$ channels
including the $3p\,4l'$ resonances. Despite the convolution of the DR
cross section with the plasma electron energy distribution one can
recognize that the calculations of \citet{Gu2004a} and
\citet{Altun2006a} either underestimate the strengths of the
$3p\,4l'$ resonances at low energies or overestimate their energy
positions. We can reproduce the qualitative trend of their plasma DR
rate coefficients by shifting our experimentally derived $3p\,4l'$
resonances  to higher energies by about $0.15~\text{eV}$. Both
theoretical plasma DR rate coefficients underestimate the
experimentally-derived rate coefficient in the photoionized zone. The
rate coefficient of \citet{Altun2006a} is a factor of $0.71$--$0.86$
lower than the experimentally derived rate coefficient. For the rate
coefficient of \citet{Gu2004a} the factor is $0.83$--$1.02$.

The comparison of our experimentally derived plasma DR rate
coefficient with the experimental result of \citet{Orban2006a} shows,
that both data sets agree to within $25\%$ in the temperature rage
$0.1$--$1000~\text{eV}$. In the photoionized zone the plasma DR rate
coefficient of \citet{Orban2006a} is $18\%$--$24\%$ larger than our
result. This larger deviation than in the collisionally ionized zone
is probably associated with differences in the data reduction
process. In the collisionally ionized zone the plasma DR rate
coefficient of \citet{Orban2006a} is less than $5\%$ lower than our
result. The increasing deviation between the two plasma DR rate
coefficients above $\approx30~\text{eV}$ is probably attributed to
the fact that, in contrast to the work of \citet{Orban2006a}, the
present experimentally derived DR rate coefficient also includes
$\Delta N=1$ and even $\Delta N>1$ DR.

\section{Conclusions}\label{sec::conclusions}

Electron-ion recombination of \ion{Si}{4} forming \ion{Si}{3} was
studied both experimentally using the merged-beams method at a heavy
ion storage ring and theoretically by employing the MCDF method. We
see good agreement in DR resonance strength and positions between the
experiment and the MCDF calculations for $\Delta N=0$ DR in the
investigated electron-ion collision energy range $0$--$6~\text{eV}$.
Below an energy of $1.4~\text{eV}$ the accuracy in the energy
positions was better than $70~\text{meV}$, the position of the
resonance at the lowest energy was even accurate to within
$20~\text{meV}$. A great advantage of the MCDF method is that it is
conceptually much simpler to implement than other many--body
techniques and, hence, can be applied also to more complex shell
structures --- if enough computational resources are available. We
currently plan to extend the code in order to make computations
feasible for atoms and ions with (initially) two or three electrons
in their valence shell.

The present experimentally derived \ion{Si}{4} plasma DR rate
coefficient agrees with the experimental result of
\citet{Orban2006a}, to within the combined experimental errors. We
found good agreement between the theoretical results of
\citet{Gu2004a} and \citet{Altun2006a} and our experimental result in
the temperature range where \ion{Si}{4} forms in collisionally
ionized plasmas. The agreement is reasonable at temperatures where
\ion{Si}{4} is predicted to form in photoionized as. At temperatures
below this the agreement becomes significantly worse with decreasing
temperature. These findings demonstrate the necessity of benchmarking
theoretical results with experiment, because modern theory still has
difficulty calculating resonance energies reliably when the
electron-ion collision energy and Rydberg level of the recombined ion
are small.

\begin{acknowledgements}
We gratefully acknowledge the excellent support by the MPI-K
accelerator and TSR crews. This work was supported in part by the
German federal research-funding agency DFG under contract no.\
Schi~378/5 and Fr~1251/13. DL and DWS were supported in part by the
NASA Astronomy and Astrophysics Research and Analysis program and the
NASA Solar and Heliospheric Physics program.
\end{acknowledgements}


\end{document}